 \documentclass[preprint2,longabstract]{emulateapj}

\newcommand{\oi}{[O~{\sc i}]}
\newcommand{\neii}{[Ne~{\sc ii}]}
  
\newcommand{\Ha}{H$\alpha$}

\newcommand{\Rsun}{R$_\odot$}
\newcommand{\Lsun}{L$_\odot$}
\newcommand{\Msun}{M$_\odot$}
\newcommand{\Lacc}{$L_{acc}$}
\newcommand{\Macc}{$\dot M_{acc}$}

\slugcomment{Not to appear in Nonlearned J., 45.}

\shorttitle{The origin of the {\oi} LVC}
\shortauthors{Rigliaco et al.}

\begin{document}


%\title{THE ORIGIN OF THE {\oi} LOW-VELOCITY COMPONENT\\ FROM YOUNG STELLAR OBJECTS}
\title{UNDERSTANDING THE ORIGIN OF THE \\
{\oi} LOW-VELOCITY COMPONENT FROM T TAURI STARS 
% HINTS OF AN FUV-DRIVEN PHOTOEVAPORATIVE WIND
}


\author{E. Rigliaco\altaffilmark{1},  I. Pascucci\altaffilmark{1}, U. Gorti\altaffilmark{2,3}, S. Edwards\altaffilmark{4} and D. Hollenbach\altaffilmark{2}}
\affil{${^1}$Lunar and Planetary Laboratory, The University of Arizona, Tucson, AZ 85721, USA}
\email{rigliaco@lpl.arizona.edu}

%\author{U. Gorti\altaffilmark{3} and D. Hollenbach\altaffilmark{3}}
\affil{${^2}$SETI Institute, Mountain View, CA, USA}
\affil{${^3}$NASA Ames Research Center, Moffett Feld, CA}

%\author{S. Edwards\altaffilmark{2}}
\affil{${^4}$Astronomy Department, Smith College, Northampton, MA 01063}

%\and



\begin{abstract} 
The formation time, masses, and location of planets are strongly impacted by the physical mechanisms that disperse protoplanetary disks and the timescale over which protoplanetary material is cleared out. 
Accretion of matter onto the central star, protostellar winds/jets, magnetic disk winds, and photoevaporative winds operate concurrently. 
Hence, disentangling their relative contribution to disk dispersal requires identifying diagnostics that trace different star-disk environments. 
Here, we analyze the low velocity component (LVC) of the Oxygen %{O~{\sc i}} 
optical forbidden lines, which is found to be blueshifted by a few km/s with respect to the stellar velocity. 
We find that the {\oi} LVC profiles are different from those of {\neii} at 12.81$\mu$m and CO at 4.7$\mu$m lines pointing to different origins for these gas lines. 
We report a correlation between the luminosity of the {\oi} LVC and the accretion luminosity \Lacc. 
We do not find any correlation with the X-ray luminosity, while we find that the higher is the stellar FUV luminosity, the higher is the luminosity of the {\oi} LVC.
In addition, we show that the {\oi}6300\AA/5577\AA\ ratio is low (ranging between 1 and 8). 
These findings favor an origin of the {\oi} LVC in a region where
OH is photodissociated by stellar FUV photons and argue against thermal emission from an
X-ray-heated layer.
Detailed modeling of two spectra with the highest S/N and resolution shows that there are two components within the LVC: a broad, centrally peaked component that can be attributed to gas arising in a warm disk surface in Keplerian rotation (with FWHM between $\sim$40 and $\sim$60~km/s), and a narrow component (with FWHM$\sim$10~km/s and small blueshifts of $\sim$2~km/s)  that may arise in a 
cool ($\lesssim$1,000~K) molecular wind.
\end{abstract}



\section{Introduction} \label{intro}

Planets form in circumstellar disks of gas and dust around young stars. 
Hence, understanding how protoplanetary disks evolve and disperse is essential to our understanding of the planet formation process. 
Various mechanisms have been proposed to explain how disks evolve and disperse (see e.g. Hollenbach et al. 2000 for a review): viscous accretion (e.g. Lynden-Bell \& Pringle 1974); MHD winds arising over a narrow range of disk radii near the corotation radius (X-winds, e.g. Shu et al. 2000); MHD winds originating over a large range of disk radii both inside and outside the corotation radius 
(magnetocentrifugal disk winds, e.g. Koenigl \& Pudritz 2000); photoevaporative winds driven by high-energy photons from the central star (e.g. Clarke et al. 2001), and planet formation (e.g. Hillenbrand 2008).
Determining which of these mechanisms dominates requires identifying diagnostics tracing these processes as disks evolve. 

High-resolution optical spectroscopy of T~Tauri stars (hereafter TTs) has been crucial in identifying accretion and wind diagnostics. For instance, it is now well established that broad (several hundred km/s) permitted Hydrogen lines mostly trace disk gas falling onto the star in magnetospheric accretion columns (e.g. Calvet et al. 2000 for a review). Forbidden emission lines (such as the bright {\oi} line at 6300\,\AA) typically exhibit two distinct components: a high-velocity component (HVC) blueshifted by 50--200~km/s with respect to the stellar velocity, and a low-velocity component (LVC), typically blueshifted  by only $\sim$5-20km/s (Hartigan et al. 1995, hereafter HEG, Panoglou et al. 2012). The HVC is unambiguously tracing collimated jets that in some cases have been spatially resolved at distances from 50 to several AU from the star (e.g. Hirth et al. 1997, Hartigan et al. 2004).

The origin of the LVC is uncertain. The fact that the line ratios in the LVC are very different from those in the HVC indicates that the two components arise from physically distinct regions (Edwards et al. 1989, Hirth 1994, Hamann 1994, HEG). In addition, the finding that the LVC peak velocities  are ordered inversely with their respective critical densities (e.g. $\mid$v$_{([SII]6731\AA)}\mid$~$\ge$~$\mid$~v$_{([OI]6300\AA)}\mid$) hints at a disk wind origin (HEG). In a disk wind, lines with higher critical densities (like {\oi}) form closer to the disk than lines with lower critical densities (like [S~{\sc ii}]) and the flow accelerates as it rises from the disk (HEG).  
However, MHD winds  are generally launched at speeds greater than the local Keplerian velocity (e.g., K\"onigl and Salmeron 2011), and the typical radii at which the MHD winds originate are too small to be compatible with the observed small blueshifts in the LVC (see Ferreira et al. 2006 for a review). 
 
More recently, a number of authors have explored the possibility that the LVC of forbidden lines, and in particular the bright {\oi} 6300\,\AA\ line, trace a thermal photoevaporative wind driven by high-energy photons from the central star (e.g. Font et al. 2004, Ercolano \& Owen 2010, Gorti et al, 2011). 
These authors have considered the different types of stellar radiation that could thermally drive a photoevaporative flow and explain the properties of the {\oi} LVC. Stellar EUV   photons (13.6\,eV $< h \nu \le$\,100\,eV) can be easily ruled out, since the EUV-heated layer cannot reproduce the large {\oi} LVC luminosities that are observed (Font et al. 2004, Hollenbach \& Gorti 2009, Ercolano et al. 2009, Owen et al. 2010, Ercolano \& Owen 2010). 
This is primarily because the EUV-heated layer is mostly ionized and the fraction of neutral gas (including atomic oxygen) is very low. Neutral gas, on the other hand, needs to be relatively hot ($\sim$8,000~K) to thermally populate the upper levels of the {\oi}  transitions. Heating by hard X-rays ($h \nu >$1~keV) is insufficient as typical gas temperatures achieved are relatively low with $T\sim 1,000-4,000$K (Hollenbach \& Gorti 2009, Ercolano et al. 2009, Owen et al. 2010, Ercolano \& Owen 2010). A significant flux of soft X-rays ($\sim 0.1-0.3$keV) capable of heating gas to $\sim8,000$K is needed to produce typical {\oi} LVC luminosities. Gas at these temperatures and beyond the gravitational radius\footnote{The gravitational radius ($r_g$) is the radius where the thermal speed is equal to the escape speed from the disk.} is likely to photoevaporate and the observed {\oi} LVC has been attributed to this mechanism (Ercolano \& Owen 2010). 
A detailed study of the nearby disk of TW~Hya has raised yet another possibility: that the {\oi} emission comes from cooler gas located below a surface layer, and thus the lines trace a bound disk layer where stellar FUV photons (6\,eV $< h \nu \le$\,13.6\,eV) dissociate OH molecules (Gorti et al. 2011). In this model most of the {\oi} emission comes from OH photodissociation (a non-thermal process) and there is only a small contribution from thermal emission. The model accounts for three observational properties of the {\oi} lines detected toward TW~Hya: i) a low {\oi}6300\AA/{\oi}5577\AA\ line ratio;  ii) peak velocities centered at the stellar velocity; and iii) small FWHMs of $\sim$10\,km/s (Pascucci et al. 2011). 
Photodissociation of OH and H$_2$O molecules was also proposed by Acke et al. (2005) to explain the origin of the observed {\oi}6300\AA\ luminosity in a sample of Herbig Ae/Be stars observed with high-resolution spectroscopy.

Recently, other near- and mid-infrared lines have been proposed as wind tracers. 
Using low-resolution spectra it has been possible to claim that the {\neii} line at $\lambda$12.81$\mu$m is an important diagnostic for gas in the disk surface (e.g. G\"udel et al. 2010, and references therein). However, ground-based higher resolution observations, gave us more information on the line kinematic. Indeed, 
this line also shows  two distinct components like the optical forbidden lines. 
Similar to the {\oi} line, the HVC of the {\neii} is clearly associated with jets/outflows (it has been also spatially resolved toward one source, see van Boekel et al. 2009), while the profile and peak velocity of the LVC is consistent with a photoevaporative disk wind driven by stellar X-ray/EUV photons (Pascucci \& Sterzik  2009, Sacco et al. 2012, Baldovin-Saavedra et al. 2012). Thus, the {\neii} LVC  traces either the EUV-heated or the X-ray-heated disk layer we discussed above in the context of the {\oi} LVC (Meijerink et al. 2008, Hollenbach \& Gorti 2009, Ercolano \& Owen 2010). 
A less certain diagnostic of disk winds is the fundamental CO $v$=1--0 emission at 4.7\micron. 
Spectroastrometry in the CO line (Pontoppidan et al. 2011) shows non-Keplerian motions in a few sources. Those motions are consistent with  wide-angle disk-winds. Following up on this finding Bast et al. (2011) and Brown et al. (2013) interpret the small blueshift in the CO peaks ($\le$5\,km/s) arising from slow disk wind. 

The next step in these studies is to compare different wind diagnostics to understand the structure and evolution of disk winds. If the {\oi} lines arise from thermal emission in a mostly neutral photoevaporative wind, similarities with observed {\neii} emission may be expected if X-rays partially ionize the Neon in the mostly neutral gas. 
{\oi} produced as an OH photodissociation product will arise in mainly molecular gas and may be more correlated with CO emission. Alternately, all three diagnostics may trace different regions of the same wind that is molecular at the base, atomic in the middle, and ionized in its upper layers. 

In this paper, we compare the emission from these three proposed wind diagnostics. So far such a comparison has been carried out only for four disks and has been restricted to the {\neii} line and optical forbidden lines. 
Pascucci et al. (2011) analyzed TWHya, and they found that the {\neii} and the {\oi} lines come from two physically different regions. 
Baldovin-Saavedra et al. (2012) extended the analysis to three objects. For CoKuTau1 they found that the two lines seem to have the same origin, but the comparison does not allow to conclusively determine the origin of the lines (disk atmosphere or inner bipolar jet). In FS~TauA both the lines appear to come from a EUV/X-ray driven  photoevaporative flow. For V853~Oph they found that the {\neii} line is tracing a jet, as only the HVC is  detected, while the {\oi} lines are compatible with an origin in a photoevaporative wind. 
According to these previous studies, in three out of four objects analyzed, the {\neii} and the {\oi} lines seem to arise from physically different regions.  In this paper we expand these studies in two ways: i) by extending the comparison to the fundamental CO ro-vibrational line; and ii) by almost tripling the number of objects where different wind diagnostics are analyzed. Focusing on the {\neii}, {\oi}, and CO, we aim at tracing the ionized, atomic, and molecular layer of such winds covering disk temperatures from $\sim$10,000 to 1,000\,K. We also re-analyze already published optical spectra of 30 T Tauri stars with disks (HEG sample) to further constrain the origin of the {\oi} emission.

\section{Data Collection} 

In the following subsections we present our samples  and briefly describe the data reduction and analysis (Sections~\ref{sect_sampleI} and \ref{sect_hegsample}). We also provide homogeneously derived effective temperatures, stellar luminosities, and accretion luminosities for all targets (see Sections~\ref{sect_stell_par} and \ref{sect_accr_prop}). These quantities will be used in Sections~\ref{sect_obs_res} and \ref{obs_constr} to investigate the origin of the {\oi} LVC.
\begin{deluxetable*}{c c c c c c c}
\tablewidth{0pt}
\tabletypesize{\scriptsize}
\tablecaption{Sample I sources \label{obj_Tab}}
\tablehead{
\colhead{Star} & \colhead{RA} & \colhead{DEC} &  \colhead{Instrument} & \colhead{Spectral Coverage} & \colhead{{\neii},CO } & \colhead{Ref}\\
%\colhead{} & \colhead{} & \colhead{} & \colhead{}  & \colhead{}\\
}

\startdata
S CrA N    & 19:01:08.6 & -36:57:20.0  & UVES	& 3750 -- 6800\AA & CO & 1 \\
V866 Sco & 16:11:31.4 & -18:38:24.5 	& UVES	& 3750 -- 6800\AA & CO & 2	 \\
DR Tau    & 04:47:06.2 & +16:58:42.9 	& UVES	 & 3750 -- 7500\AA & CO & 1, 3\\
VZ Cha    & 11:09:23.8 & -76:23:20.8 	& UVES	 & 4800 -- 6800\AA & CO & 4 \\
RU Lup    & 15:56:42.30 & -37:49:15.4 	& UVES	 &  3500 -- 6800\AA & {\neii}+CO & 1, 4, 5, 6, 7, 10 \\
MP Mus   & 13:22:07.53 & -69:38:12.2 	& UVES	 & 3300 -- 6800\AA & {\neii} & 7\\
V4046 Sgr & 18:14:10.48 & -32:47:34.4 & UVES	& 3500 -- 6800\AA  & {\neii} &  8 \\
TW Hya	 & 11:01:51.91	& -34:42:17.03& FEROS	& 3500 -- 9200\AA  &  {\neii}+CO & 7, 9\\
\enddata
\tablerefs{
(1) Petrov et al. 2011;
(2) Melo, C. 2003;
(3) Fischer et al. 2011; 
(4) Stempels \& Piskunov 2003;
(5) Stempels et al. 2007; 
(6) Stempels \& Piskunov 2002;
(7) Curran et al. 2011;
(8) Stempels \& Gahm, 2004;
(9) Alencar \& Batalha 2002; 
(10) Takami et al. 2001
}
\end{deluxetable*}

\subsection{Sample I}
\label{sect_sampleI}
We have assembled a sample of eight TTs with archival and literature high-resolution optical spectra and either high-resolution N-band (covering the {\neii} at 12.81\micron) or  M-band (covering the CO ro-vibrational band at 4.7\micron) spectra. We took into account the sources presented in Pascucci \& Sterzik (2009), Sacco et al. (2012) and Bast et al. (2011) and selected those whose LVC in either the {\neii} or CO lines hints at a slow disk wind  (blueshifts in the line of a few up to $\sim$10~km/s). 
Table~\ref{obj_Tab} lists the instruments used to acquire the optical spectra, the wavelength coverage, references related to the observing programs, as well as a designation to show if there are {\neii} and/or CO spectra available. 
Four out of the eight objects reported in Table~\ref{obj_Tab} (S~CrA~N, V866~Sco, DR~Tau and VZ~Cha) have been observed in M-band by Bast et al. (2011). Two objects (MP~Mus and V4046~Sgr) have been observed in N-band (Sacco et al. 2012), and the remaining two objects (RU~Lup and TW~Hya) have been observed both in M- (Bast et al. 2011) and in N-band (Sacco et al. 2012, Pascucci et al. 2011). As we will see in Sect.~\ref{sect_accr_prop}, except for TW~Hya and VZ~Cha, the objects observed in M-band are the ones with higher accretion rates.
 
The optical spectrographs used to observe the targets are VLT/UVES and VLT/FEROS. 
VLT/UVES was used with different setups, mean resolution R$\sim$70,000 
($\Delta v$=4.3 km/s), and with slit width between 0.6$^{\prime\prime}$ and 1.0$^{\prime\prime}$ depending on the observing programs.  
FEROS is a fiber-fed spectrograph with two fibers, one positioned on the star and the other on the sky (the fiber diameter projected on the sky is 2.7$^{\prime\prime}$) from which the sky background is estimated. The mean resolution of the spectrograph is R$\sim$48,000 
($\Delta v$=6.0 km/s).  
The FEROS spectra are from Alencar \& Batalha (2002).

We reduced the archival VLT/UVES spectra with the ESO/UVES Common Pipeline Library version 3.9.0 which includes bias subtraction, flat fielding, wavelength calibration, and optimal spectrum extraction. 
To  analyze the line profiles we removed contaminations due to photospheric and terrestrial lines.  
Following standard procedures (see e.g. HEG), we first subtracted the terrestrial night-sky emission lines. 
Then we subtracted telluric absorption lines using telluric standard stars (O-B spectral type) at the same resolution as the target star. 
Finally, the photospheric absorption lines were subtracted from the target spectra using the spectra of stars without disks that match the spectral type of the target star, and with known rotational velocity. In this last step we broadened the template spectra to the same rotational velocity of the target star to properly subtract photospheric lines. 
To define the stellar velocity we have used the photospheric Ca~{\sc i} line at 6439\AA. 
An example of our reduction steps applied to the spectrum of V866~Sco is shown in the Appendix (Fig.~\ref{v866_redu}).

\subsection{Sample II}
\label{sect_hegsample}
With the aim of testing whether there is any correlation between the LVC of the {\oi} lines and stellar 
properties,  we also re-analyzed already published high-resolution optical spectra from the comprehensive T Tauri survey of HEG. 
This survey includes 42 sources: ten stars are non-accreting weak TTs (WTTs) while 32 are accreting classical TTs (CTTs). In this sample we consider the 30 CTTs that have published H$\alpha$ profiles (Beristain et al. 2001), and we exclude DR~Tau, which is included in Sample I. 

The spectra were collected with the 4\,m  Mayall telescope at the Kitt Peak National Observatory between 1988 and 1992. 
The wavelength coverage is 5,000--6,800 \AA\  in the red and  4,000--4,950 \AA\ in the blue,  with a resolution of $\Delta v\sim$12\,km/s, and slit width 1.25$^{\prime\prime}$. 
Details on the data reduction can be found in HEG. 
The data analysis and the measurements of the equivalent widths (EWs) was done by HEG
on the residual line profiles, after subtracting from the target spectra an artificially veiled photosphere of a star with no disk and the same spectral type as the target star, and after shifting the spectra to the rest velocity of the stellar photosphere. 
We note that the Sample II includes objects spanning over all the accretion luminosity range for solar-mass T Tauri stars (see Sect.~\ref{sect_accr_prop}). 

\subsection{Stellar parameters}
\label{sect_stell_par}
Stellar parameters for the Sample I and Sample II sources
are listed in Table~\ref{param_Table} and ~\ref{param_HEG}, respectively. 
 
Effective temperatures and stellar luminosities for the Sample I sources were computed using spectral types (SpT), I-, R- or V-magnitudes, and extinctions found in the literature.  
In particular, stellar luminosities have been computed from the dereddened I-band magnitude and the bolometric correction corresponding to the assigned SpT (as reported by Luhman et al. 1999). When the I-band magnitude was not available, we used in order R- or V-band magnitudes, taking into account the corresponding extinction law (Mathis 1990). 
The uncertainty in the SpT of about half a subclass translates into a +/- 70\,K uncertainty on the stellar effective temperature. 
Uncertainties in the stellar luminosity ($L_*$) depend on: the adopted bolometric correction/temperature; I-, V- or R-magnitude values (known to within $\sim$0.2\,mag); the error on the extinction, which is $\sim$20\% in a typical TTs (Hartigan et al. 1991), and the distance of the source. 
We estimate a typical uncertainty in the luminosity of $\sim$0.5 dex.   
Stellar masses ($M_*$) are estimated from the location of the sources in the HR diagram using Siess et al. (2000) evolutionary tracks.
Stellar radii are estimated from stellar luminosities and effective temperatures.  
The determination of stellar masses and radii are also affected
by the distance uncertainty, typical errors on these parameters are $\sim$0.05 dex.

For the Sample II sources, we have used the luminosities and the effective temperatures 
reported in HEG, but we have recomputed the stellar masses  using the evolutionary tracks of Siess et al. (2000) for consistency with the Sample I sources (see  Table~\ref{param_HEG}).  
Tables~\ref{param_Table}--\ref{param_HEG}, we have marked with a single asterisk the multiple systems, and with a double asterisk the systems where two components have a separation smaller than $\sim$0.6$^{\prime\prime}$ (half of the slit width employed by HEG, see refereces to Tables~\ref{param_Table}--\ref{param_HEG}). 
We note that only five out of the total 38 objects (Sample I + Sample II) are binaries closer than  $\sim$0.6$^{\prime\prime}$. 

\begin{deluxetable*}{c c c c c c c c c c c c c} 
\tablewidth{0pt}
\tabletypesize{\scriptsize}
\tablecaption{Stellar
        Parameters for the Sample I sources\label{param_Table}}
\tablehead{
\colhead{Star}
        & \colhead{Log$L$} & \colhead{LogT} & \colhead{R$_*$}
        & \colhead{M$_*$} & \colhead{\Ha\ ew} &
        \colhead{Log\Lacc} & \colhead{V$_{mag}$} & \colhead{A$_V$} & LogL$^a_{\rm{[OI]}LVC}$ &
        \colhead{distance} & \colhead{Ref} \\
\colhead{}
        & \colhead{(\Lsun)} & \colhead{(K)} & \colhead{(\Rsun)} 
	& \colhead{(\Msun)} & \colhead{(\AA)} &
        \colhead{(\Lsun)} & \colhead{} &  \colhead{} & \colhead{(\Lsun)} &
	\colhead{(pc)} & \colhead{} \\
}
\startdata
S CrA N$^*$
        & -0.036 & 3.681 & 1.38 & 1.20 & 80.0 &
        -0.08 & 11.59 & 1.0 & -4.05 & 130 & 1 \\
V866 Sco N$^*$
        & 0.689 & 3.648 & 3.71 & 1.35 & 103.1
        &0.89 & 12.05 & 2.9 & -3.37 & 125 & 1 \\
DR Tau
        & 0.063 & 3.608 & 2.17 & 0.75 & 78.7
        &0.45 & 11.43 & 1.6 & -4.15 & 140 & 2 \\
VZ Cha
        & -0.133 & 3.623 & 1.61 & 0.95 & 40.0
        &-1.10 & 12.94 & 1.9 & -5.23 & 103 & 3 \\
RU Lup
        & 0.076 & 3.602 & 2.27 & 0.69 & 100.3
        &-0.29 & 11.44 & 0.1 & -4.47 & 140 & 4 \\
MP Mus
        & 0.057 & 3.702 & 1.40 & 1.23 & 19.5
        &-1.34 & 10.44 & 0.17 & -4.96 & 86 & 5 \\
V4046 Sgr$^{**}$
	& -0.281 & 3.628 & 1.33 & 0.99 & 42.0
        &-1.30 & 10.68 & 0.0 & -5.55 & 73 & 6 \\
TW Hya
        & -0.601 & 3.591 & 1.09 & 0.64 & 185.8 &
        -1.15 & 11.27 & 0.0 & -5.32 & 51 & 7 \\
\enddata
\tablenotetext{a}{{\oi} LVC 6300\AA\ luminosity. $^*$Binary systems with separation $\sim$1.4$^{\prime\prime}$ (Ghez et al. 1993, Ghez et al. 1997b, Reipurth \& Zinnecker, 1993) $^{**}$Spectroscopic binary (Stempels et al. 2004).}
\tablerefs{
(1)
        Prato et al. 2003;
(2)
        Muzerolle et al. 2003;
(3)
        Luhman 2007; 
(4)
        Herczeg et al. 2005;
(5)
        Cortes et al. 2009; 
(6)
        Donati et al. 2011;
(7)
        Alencar \& Batalha(2002).
}
\end{deluxetable*}

\begin{deluxetable*}{c c c c c c c c c c c c} 
\tabletypesize{\scriptsize}
\tablewidth{0pt}
\tablecaption{Stellar Parameters for the Sample II sources\label{param_HEG}}
\tablehead{
\colhead{ID} & \colhead{Star} &  \colhead{LogL$^a$} &  \colhead{LogT$^a$} & \colhead{R$_*^b$} & \colhead{M$_*^b$} & \colhead{\Ha\ ew$^c$} & \colhead{Log\Lacc} & \colhead{V$_{mag}^{a,d}$} & \colhead{A$_V^a$} & LogL$^a_{\rm{[OI]}LVC}$ & LogL$_X$ \\
\colhead{} & \colhead{} & \colhead{(\Lsun)} &  \colhead{(K)}  & \colhead{(\Rsun)} & \colhead{(\Msun)} & \colhead{(\AA)} & \colhead{(\Lsun)} & \colhead{} &\colhead{} & \colhead{(\Lsun)} & \colhead{(\Lsun)} \\
}
\startdata
  1 &   AA Tau  &  -0.190        &  3.58      &   1.73     &     0.53      &    76.9   &     -0.65
	& 11.66 & 1.3 & -3.94 & -3.39$^e$ \\
  2 &  AS 353A$^*$ &       0.570     &    3.643  &       3.15  &        1.22   &       54.6   &     -0.10  
	& 10.36 & 2.1 & -3.59 & --\\
  3& BP Tau   &       -0.060 &        3.602 &        1.71  &       0.70   &         49.0   &     -0.58 
	& 11.06 & 1.0 & -4.80 & -3.43$^e$\\
  4 &   CI Tau   &      0.020  &       3.602  &       2.10   &      0.69    &      70.6   &    -0.35 
	& 11.07 & 2.3 & -4.61 & -4.23$^f$\\
  5&   CW Tau &          0.332   &      3.672   &      2.12    &     1.60     &    276.7    &     1.79
	& 8.96 & 3.4 & -2.69 & -3.17$^f$\\
  6&   CY Tau  &           -0.330    &     3.602    &     1.56     &    0.71      &    38.9     &   -2.04 
	& 13.25 & 0.1 & -5.21 & -4.08$^e$ \\
  7&   DD Tau$^{**}$  &         -0.260    &     3.498    &     2.41     &    0.23      &   145.1    &   -0.56 
	& 12.21 & 1.9 & -3.89 & -4.06$^f$\\
  8&   DE  Tau &     0.030  &       3.533  &       2.87   &      0.32    &      71.2   &    -0.57 
	& 11.45 & 1.5 & -4.71 & -4.45$^e$\\
  9&   DF  Tau$^{**}$ &         0.310   &      3.544   &       3.72     &    0.36      &    71.9     &   -0.16 
	& 10.78 & 1.3& -3.73 & $<$-4.96$^f$\\
  10&   DG Tau  &          0.240     &    3.643     &     3.03      &   1.19        &  71.2     &   0.90 
	& 8.97 & 3.2 & -2.63 & --\\
   11 &  DK Tau$^*$  &          0.230   &      3.602   &      2.02    &      0.69     &     11.5    &    -1.15 
	& 10.45 &2.0 & -4.17 & -3.74$^f$\\
   12 &  DL Tau  &           -0.170    &      3.58     &    1.84      &    0.53       &  121.7     &  -0.16 
	& 11.35 & 1.7 & -4.70 & --\\
   13 &  DN Tau  &       0.030  &       3.602  &         2.10     &     0.69      &    17.8     &   -1.34 
	& 11.23 & 1.1 & -4.43 & -3.52$^e$\\
   14 &  DO Tau  &      0.050  &        3.58   &      2.31      &   0.52    &      75.6   &     0.51 
	& 9.70 & 4.6 & -3.63 & $<$-4.16$^f$\\
   15 &  DQ Tau$^{**}$  &         -0.020    &     3.602    &     1.85     &    0.69     &     61.6    &   -0.66 
	& 11.45 & 2.1 & -3.81 & --\\
   16 &  DS Tau  &          0.130     &     3.69      &    1.55     &    1.36        &  28.4     &  0.23 
	& 9.70 & 2.2 & -4.31 & --\\
   17 & FM Tau   &       -0.650  &       3.498  &       1.50  &       0.21   &       74.3   &     -1.53 
	& 13.10 & 1.2 & -5.22 & -3.52$^e$\\
   18 & FP Tau    &      -0.550  &       3.477  &       1.89  &       0.17   &       24.8   &      -2.72 
	& 13.91 & 0.0 & -5.51 & $<$-3.89$^f$\\
   19& GG Tau$^{**}$   &          0.170     &     3.58      &   2.66      &   0.52       &     45.0       &  -0.66 
	& 11.10 & 1.25 & -4.55 & $<$-4.98$^f$\\
   20& GI Tau$^*$    &      0.070  &        3.58   &      2.32  &      0.52    &      12.8   &     -1.65 
	& 11.39 & 1.7 & -4.57 & -3.66$^e$\\
   21& GK Tau$^*$   &       0.050  &       3.602  &       2.12  &       0.69   &       17.6  &      -1.56 
	& 11.59 & 0.78 & -4.70 & -3.56$^f$\\
   22& GM Aur   &       0.230   &      3.602   &      2.02   &       0.69     &    110.5   &    -0.57 
	& 11.93 & 0.10 & -4.92 & -3.75$^e$\\
   23& HK Tau$^*$   &    -0.040   &      3.602 &        1.83 &        0.69  &        49.5 &      -0.39 
	& 10.75 & 5.0 & -4.18 & -4.44$^f$\\
   24& HN Tau$^*$  &         -0.560      &   3.602    &     1.14    &     0.77     &     90.8    &   -0.94 
	& 12.33 & 1.9 & -3.71 & $<$-3.88$^e$ \\
   25& LkCa 8    &        -0.390      &    3.58     &    1.45      &   0.54      &     7.9      &  -2.79  
	& 12.79 & 0.25 & -4.92 & --\\
   26& RW Aur$^{**}$  &           0.360       &  3.662     &    2.31     &     1.51      &       70.0    &     0.94  
	& 8.89 & 2.2 & -3.65 & --\\
   27& UY Aur$^*$   &        0.120     &     3.58     &     2.45   &        0.52     &      76.7   &      -0.29  
	& 11.07& 1.3 & -3.93 & --\\
   28& UZ Tau  E$^*$&           0.440       &  3.553      &    4.22    &      0.39     &      70.4   &      0.12  
	& 10.28 & 2.8 & -3.34 & -4.74$^e$\\
   29& V836 Tau &          -0.260      &   3.602     &     1.58     &     0.70        &    2.4     &     -3.23 
	& 12.23 & 0.9 & -4.95 & $<$-3.58$^f$\\
   30& YY Ori   &       0.332     &    3.602    &      2.64    &      0.68       &    25.9   &     0.07 
	& 11.85 & 1.9 & -3.61 & --\\
\enddata
\tablenotetext{}{$^a$Stellar luminosities, effective temperatures, V-magnitude, extinctions (A$_V$), and {\oi}~LVC 6300\AA\ luminosities as published by HEG. $^b$Masses and radii have been recomputed considering the Siess et al. (2000) evolutionary tracks. $^c$\Ha\ equivalent widths as published by Beristain et al. (2001). $^d$ Dereddened V-magnitude. $^*$Binary systems with separation larger than 0.6$^{\prime\prime}$. $^{**}$Binary systems with separation smaller than 0.6$^{\prime\prime}$ (Mathieu et al. 1997, Silber et al. 2000, White \& Ghez 2001).
} 
\tablerefs{
$^e$ G\"udel et al. (2007); 
$^f$ Stelzer \& Neuh\"auser (2001). 
}
\end{deluxetable*}

\subsection{Accretion luminosities}
\label{sect_accr_prop}

Accretion luminosities (\Lacc) for the Sample II stars were first computed by HEG from the 'veiling'\footnote{the veiling is defined as $\frac{EW_{true}}{EW_{meas}}-1$ where $EW_{true}$ is the EW of the absorption line when no veiling is observed and $EW_{meas}$ is the measured EW.} of stellar absorption features at optical wavelengths, basically measuring the hot continuum excess produced by the accretion shock at the stellar surface. 
Recent studies have found that the luminosity of gas lines produced in magnetospheric accretion columns (e.g. the Hydrogen recombination lines \Ha, H$\beta$, Pa$\gamma$, Pa$\beta$ among others) correlates with the accretion luminosity (\Lacc , see e.g., Herczeg \& Hillenbrand 2008, Rigliaco et al. 2012 and references therein).  
In the wavelengths range covered by the Sample I observations, we have been able to obtain an estimate of the accretion luminosity based on the \Ha\ and H$\beta$ line luminosities. However, for the Sample II objects only the \Ha\ line profiles and EWs have been published in the literature (Beristain et al. 2001). Thus, even if the Log\Lacc--Log$L_{H\beta}$ relationship has been found to have a smaller scatter than the Log\Lacc--Log$L_{H\alpha}$ relationship (Herczeg \& Hillenbrand 2008, Fang et al. 2009, Rigliaco et al. 2012), we will use the  \Ha\ lines, obtained simultaneously with the {\oi} lines, 
to recompute accretion luminosities  consistently for both Sample I and Sample II objects.  
The errors introduced by considering the H$\alpha$ line are, qualitatively, a factor of two larger than the errors we would have on the accretion luminosities using the H$\beta$ line. 

To measure \Lacc{}  we use the most recent relation  obtained by Rigliaco et al. (2012) from a sample of low-mass stars and brown dwarfs in the Taurus-Molecular Cloud, in the young open cluster IC~348 and in the $\sigma$~Orionis cluster:
\begin{equation}
\label{eq_lacc}
Log\frac{{L}_{acc}}{L_{\odot}}  = (2.99 \pm 0.16) + (1.49 \pm 0.05)\times Log\frac{L_{H\alpha}}{L_{\odot}}
\end{equation}

In the mass range considered here, this relation is consistent with those found by other groups (Herczeg \& Hillenbrand 2008, Fang et al. 2009). To compute the H$\alpha$ luminosity ($L_{H\alpha}$) we first measure the H$\alpha$ EW by integrating below the entire line. %(see Fig.~\ref{Ha_prof}).
Then, we use literature V-band magnitudes for the continuum and the extinction values listed in Tabs.~\ref{param_Table} and \ref{param_HEG} to obtain $L_{H\alpha}$. We note that we prefer to rely on \Lacc\ rather than on mass accretion rates  because computing the latter requires determining stellar masses and radii which add additional uncertainties. Moreover, it requires making additional assumptions about poorly constrained  disk properties (see e.g. Gullbring et al.~1998).

\subsection{{\oi} line profiles characterization}
\label{ew_section}

To characterize the LVC of the {\oi} lines for the Sample I objects we compute the equivalent widths (EWs) and associated uncertainties using a Monte~Carlo approach. We add a normally distributed noise at each wavelength with an amplitude equal to the flux uncertainty at that wavelength. We then make 1,000 realizations of the spectra and EW measurements and take as resulting EW and uncertainty the mean and standard deviation of the distribution (see Pascucci et al.~2008 for further details on the procedure).   
Five out of the eight Sample I objects show a clear LVC in the {\oi}6300\AA\ line. 
Two out of eight objects (namely SCrA~N, and RU Lup\footnote{these are classified disk wind sources by Bast et al.~(2011) based on the CO v=1-0 profile}) have a HVC that strongly compromises the LVC, while in one object (V866~Sco$^3$) the LVC is slightly compromised by the HVC (we note that the HVC is always more pronounced in the {\oi}6300\AA\, than in the 5577\AA\, transition, see also HEG). 
For these three objects we compute the EW of the LVC  integrating over the portion of the spectrum between $-$60~km/s and $+$60~km/s, following the same method used by HEG. 
For all the Sample I objects we also measure line luminosities from the line EWs and the total dereddened flux, assuming the distances, magnitudes and extinctions listed in Table~\ref{param_Table}. 
Finally, we provide two additional values to characterize the line profile: the velocity at which the emission peaks ($v_{peak}$)\footnote{The profiles cannot be well reproduced by a single-gaussian, therefore we obtained the peak centroid by averaging the two consecutive velocity values where the emission peaks.},  and the width of the profile at half of the peak emission (FWHM, see Table~4). 

Table~\ref{oi_Table} summarizes our measurements. From this Table we omit the {\oi}6363\AA\ EWs because they are always one-third of the {\oi}6300\AA\ EWs, as expected from the Einstein coefficients and energy levels. 
When the {\oi}5577\AA\, is also detected (in half of the Sample I objects), and the HVC does not contaminate the LVC at low radial velocities,  its FWHM and peak centroid is similar as that of the {\oi}6300\AA\ line. 
As such, we report only the FWHM and centroid of the {\oi} 6300\AA\ line in Table~\ref{oi_Table} for the objects with a LVC well separated from the HVC. In the other cases (SCrA~N, RU Lup, and V866~Sco) we report the FWHM and centroid of the {\oi} 5577\AA\ line.

For the Sample II objects, we use the luminosities of the {\oi}6300\AA\ line obtained by HEG (Table~5 in their paper, and second-last column of Table~3 of this paper).  
We use the HEG subsample of Sample II objects where the HVC and the LVC components can be well separated (as defined by HEG, and as reported in their Table~8). We report this subsample of objects in Table~\ref{Sample_prop_tab}. From the original salection done by HEG we have excluded DGTau, because in this object the {\oi} LVC is clearly affected by the presence of the HVC, and DRTau, which is included in our Sample I sources.  For these 12 objects, we report in Table~\ref{Sample_prop_tab} the peak velocities as reported in Table~8 of HEG, and we computed the FWHM with the same method described above.  We tested that our procedure of measuring $v_{peak}$ is consistent with the one used by HEG.

\section{Observational Results}
\label{sect_obs_res}
In the following, we briefly describe our main observational results which stem from the comparison of {\oi} line profiles to those of {\neii} and CO profiles (Sample I sources, Sect.~\ref{sect_line_prof}) and from trends between the {\oi} line luminosities, FWHMs, and peak centroids with stellar parameters (Sample I and II sources, Sect.~\ref{sect_IandII}).

\begin{figure*}
\centering
\includegraphics[scale=0.5, angle=-90]{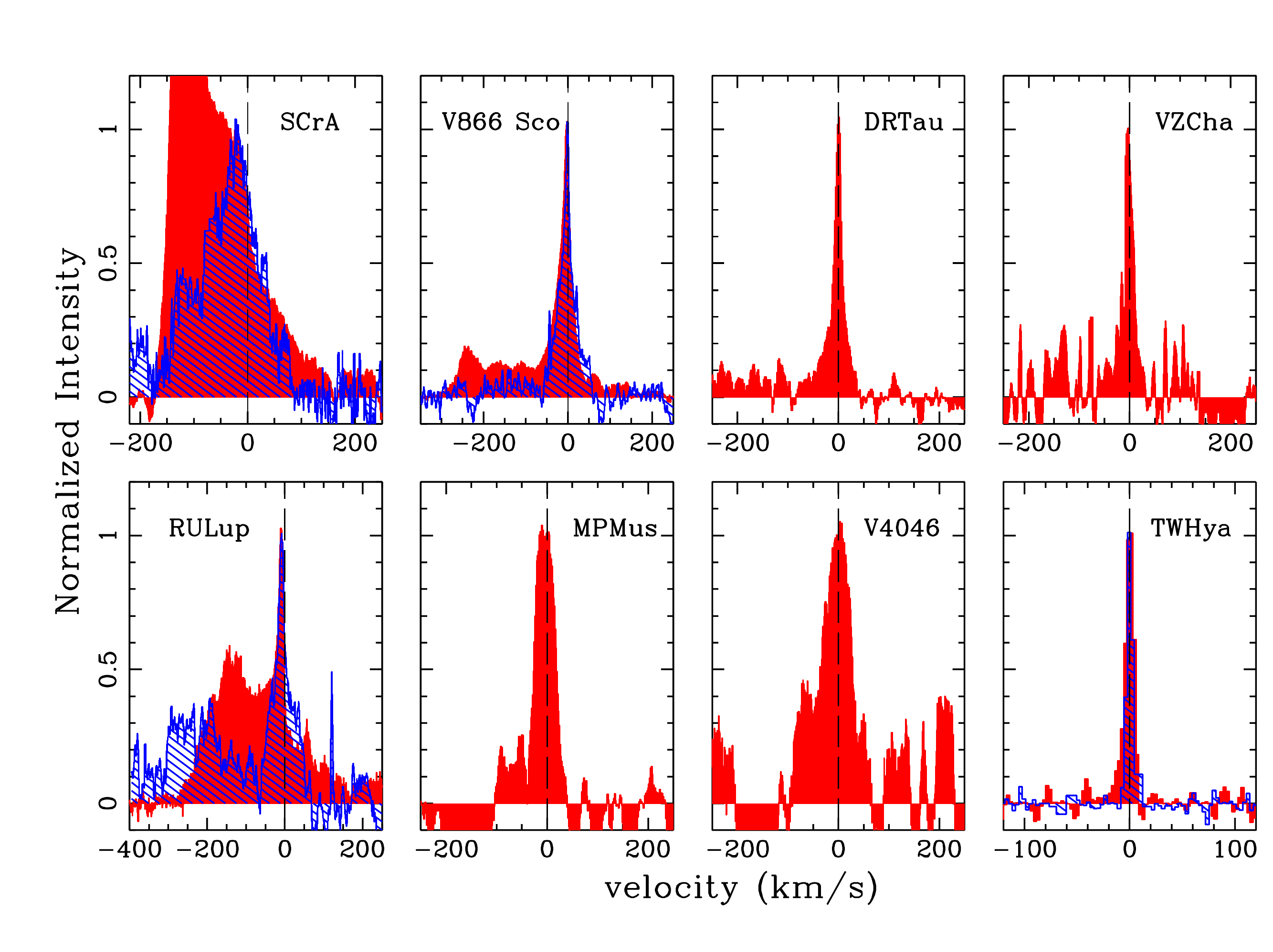}
\caption{{\small {\oi} line profiles comparison. The red area represents the profile of the {\oi} at 6300\AA\  while the blue shaded area is the 
{\oi} 5577\AA\ line profile. Profiles have been scaled to the peak of the {\oi}6300\AA\ LVC to 
highlight any difference in the shifts and/or shapes of the two lines. Note that the {\oi} 5577\AA\ line is not detected (hence not shown) toward DR~Tau, VZ~Cha, MP~Mus, and V4046~Sgr. Also note that the SCrA~N profile is has been  scaled to the peak of the {\oi}5577\AA\ line.
 }
 \label{OIsameProfile}
}
\end{figure*}

\begin{deluxetable*}{c
      c c c c c c c c} 
\tablewidth{0pt}
    
\tabletypesize{\scriptsize}
\tablecaption{Oxygen lines parameters using a gaussian fit to the LVC.\label{oi_Table}}
\tablehead{
\colhead{Star}
      & \colhead{Sample} &\colhead{EW$_{{[O{\sc I}]}6300}$}
      & \colhead{EW$_{{[O{\sc I}]}5577}$} & \colhead{FWHM$^a$} & \colhead{$v_{peak}^{a,b}$} & \colhead{$i$}
      & \colhead{Ref$^c$} \\
\colhead{}
      & \colhead{} & \colhead{(\AA)} & \colhead{(\AA)} &
      \colhead{(km/s)} & \colhead{(km/s)} &
      \colhead{($^{\circ}$)} &\colhead{} \\
}
\startdata
S CrA N$^*$
      & I & 0.83 (0.09) & 0.15 (0.04) &
      105.5$\pm8.8$ & -22.3 $\pm$ 2.0 & -- & --\\
V866 Sco$^*$
      & I & 1.13 (0.10) & 0.26 (0.05) &
      20.2$\pm0.7$ & -2.1 $\pm$ 0.7 &25 & 1\\
DR Tau
      & I & 0.28 (0.06) & $<$0.04 &
      14.1$\pm2.7$ & -0.2 $\pm$ 0.7 & 37 & 2\\
VZ Cha
      & I & 0.13 (0.04) & $<$0.04 &
      22.0$\pm6.2$ & -3.5 $\pm$ 1.5 & -- & --\\
RU Lup$^*$
      & I & 0.54 (0.08) & 0.10 (0.06) &
      21.3$\pm15.8$ & -6.4 $\pm$ 1.6 & 10 & 3 \\
MP Mus
      & I & 0.17 (0.03) & $<$0.04 &
      42.6$\pm13.6$ & -6.8 $\pm$ 2.0 & 30 & 4\\
V4046 Sgr
      & I & 0.09 (0.04) & $<$0.03 &
      47.1$\pm8.4$ & -0.6 $\pm$ 2.0 & 35 & 5\\
TW Hya
      & I & 0.53 (0.06) & 0.09 (0.05) &
      8.2$\pm0.6$ & -0.0 $\pm$ 0.7 & 7 & 6\\
\hline
AS 353A
      & II &0.39 (0.02) & $<$0.12 &
      42.8$^{+15.0}_{-25.0}$ & -7.0$^d$  & 60 & 7 \\
CW Tau
      & II & 4.3 (0.07) & 1.37 (0.06) &
      44.1$\pm1.5$ & -6.0$^d$  & 80 & 8\\
CY Tau
      & II & 0.37 (0.05) & 0.09 (0.04) &
      40.5$\pm6.6$ & -3.0$^d$  & 52 & 9\\
DF Tau
      & II & 1.27 (0.04) & 0.33 (0.03) &
      34.8$\pm2.0$ & -4.0$^d$  & 63 & 10\\
DN Tau
      & II & 0.43 (0.03) & 0.06 (0.03) &
      55.7$\pm6.3$ & -9.0$^d$  & 78 & 9\\
DQ Tau
      & II & 2.17 (0.04) & 0.38 (0.03) &
      45.1$\pm1.1$ & -13.0$^d$  & 35 & 11 \\
FM Tau
      & II & 0.45 (0.05) & 0.34 (0.05) &
      57.6$\pm6.9$ & -3.0$^d$  & 70 & 12\\
GG Tau
      & II & 0.28 (0.02) & 0.05 (0.03) &
      28.3$\pm7.0$ & -3.0$^d$  & 37 & 13\\
GK Tau
      & II & 0.31 (0.03) & 0.10 (0.04) &
      51.1$\pm5.6$ & -4.0$^d$  & 23 & 14\\
GM Aur
      & II & 0.25 (0.04) & 0.04 (0.04) &
      28.9$\pm8.5$ & +1.0$^d$  & 45 & 9\\
UY Aur
      & II & 1.16 (0.02) & 0.13 (0.02) &
      38.8$\pm5.9$ & -6.0$^d$  & 42 & 15\\
YY Ori
      & II & 0.57 (0.08) & 0.20 (0.04) &
      59.2$\pm6.0$ & -15.0$^d$  & -- & --\\
\enddata
\tablenotetext{}{$^a$FWHMs and peak velocities (in the stellocentric reference frame) are from fits to the {\oi}6300\,\AA{} line except for S~CrA~N, V866~Sco, and RU~Lup (stars with $^*$) where they refer to the  {\oi}5577\,\AA{} line. This choice is motivated by the contamination of the HVC in the {\oi}6300\,\AA{} LVC (see Fig.~1). 
{$^b$}The error on the peak centroids for the Sample II objects is $\pm$2~km/s, as reported by HEG. 
{$^c$}Ref is for the disk inclination. {$^d$} peak velocities for the Sample II objects are from HEG.}
\tablerefs{
(1)
      Andrews et al. (2010)
(2) Isella
      et al. (2009)
(3)
      Stempels \& Pistunov (2002)
(4)
      Kastner et al. (2010)
(5)
      Kastner et al. (2008)
(6) Qi et
      al. (2004)
(7) Davis
      et al. (1996);
(8) White
      \& Ghez (2001);
(9)
      Kitamura et al. (2002); 
(10) Ghez
      et al. (1997a);
(11)
      Guilloteau et al. (2011)
(12)
      Furlan et al. (2006); 
(13) Pietu
      et al. (2011);
(14)
      Weaver (1986);
(15) Close
      et al. (1997).
}
\label{Sample_prop_tab}
\end{deluxetable*}

\subsection{Sample I: Line Profile Comparison} 
\label{sect_line_prof} 
To investigate the origin of the {\oi} LVC we start by analyzing its line profile.
A visual comparison of the {\oi}6300\AA\ and {\oi}5577\AA\ lines is provided in Figure~\ref{OIsameProfile}. 
In three out the four objects where both the {\oi} lines are detected (V866~Sco, DR~Tau, and TW~Hya) this comparison  
shows that the LVC of the {\oi}6300\AA\ and {\oi}5577\AA\ lines  have similar profiles suggesting that they trace gas with analogous properties (temperature and density), and likely the same region around the star.
A similar result can be inferred from the subsample of HEG sources with distinct LVC and HVC components (see their Table~8 and the peak centroids reported for CW~Tau, DF~Tau, DR~Tau, and UY~Aur).   

\begin{figure*}
\centering
\includegraphics[scale=0.4]{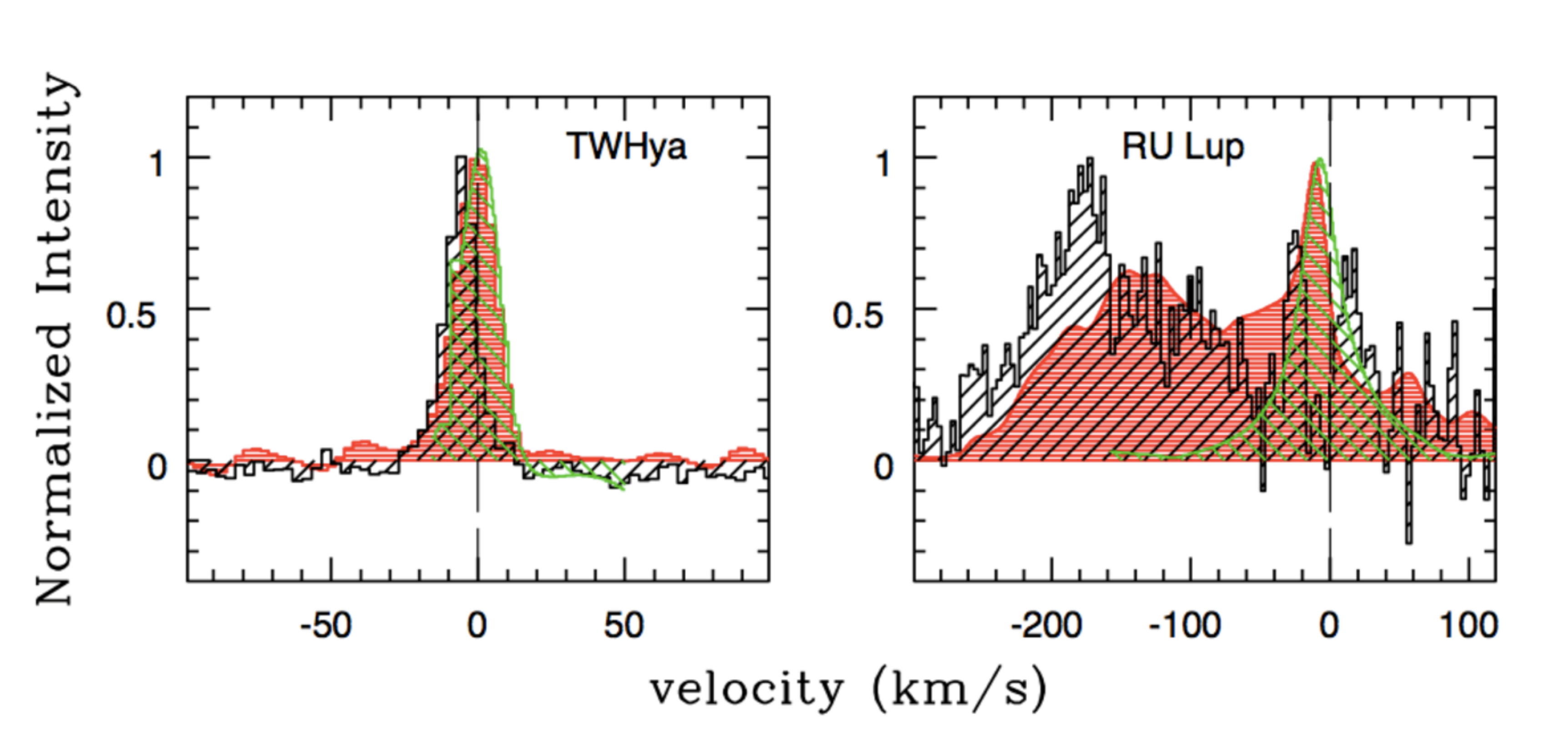}
\caption{{\small CO (green) -- {\neii} (black) -- {\oi}6300\AA\ (red) line profiles comparison. }
 \label{Ne_CO_OI_comp}
}
\end{figure*}

\begin{figure*}
\centering
\includegraphics[scale=0.40]{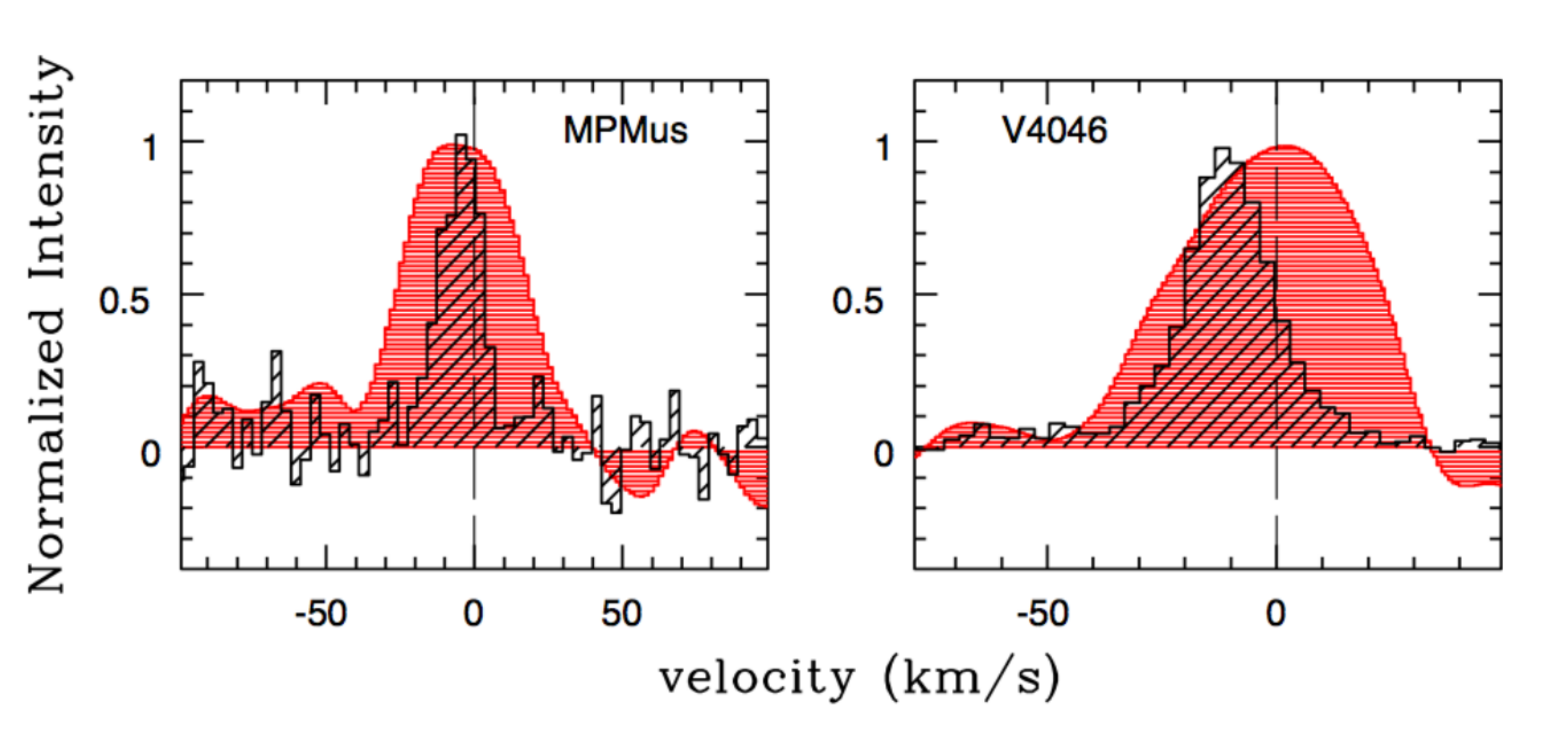}
\caption{{\small {\neii} (black) -- {\oi}6300\AA\ (red) line profiles comparison.}
 \label{Ne_OI_comp}
}
\end{figure*}

\begin{figure*}
\centering
\includegraphics[scale=0.7]{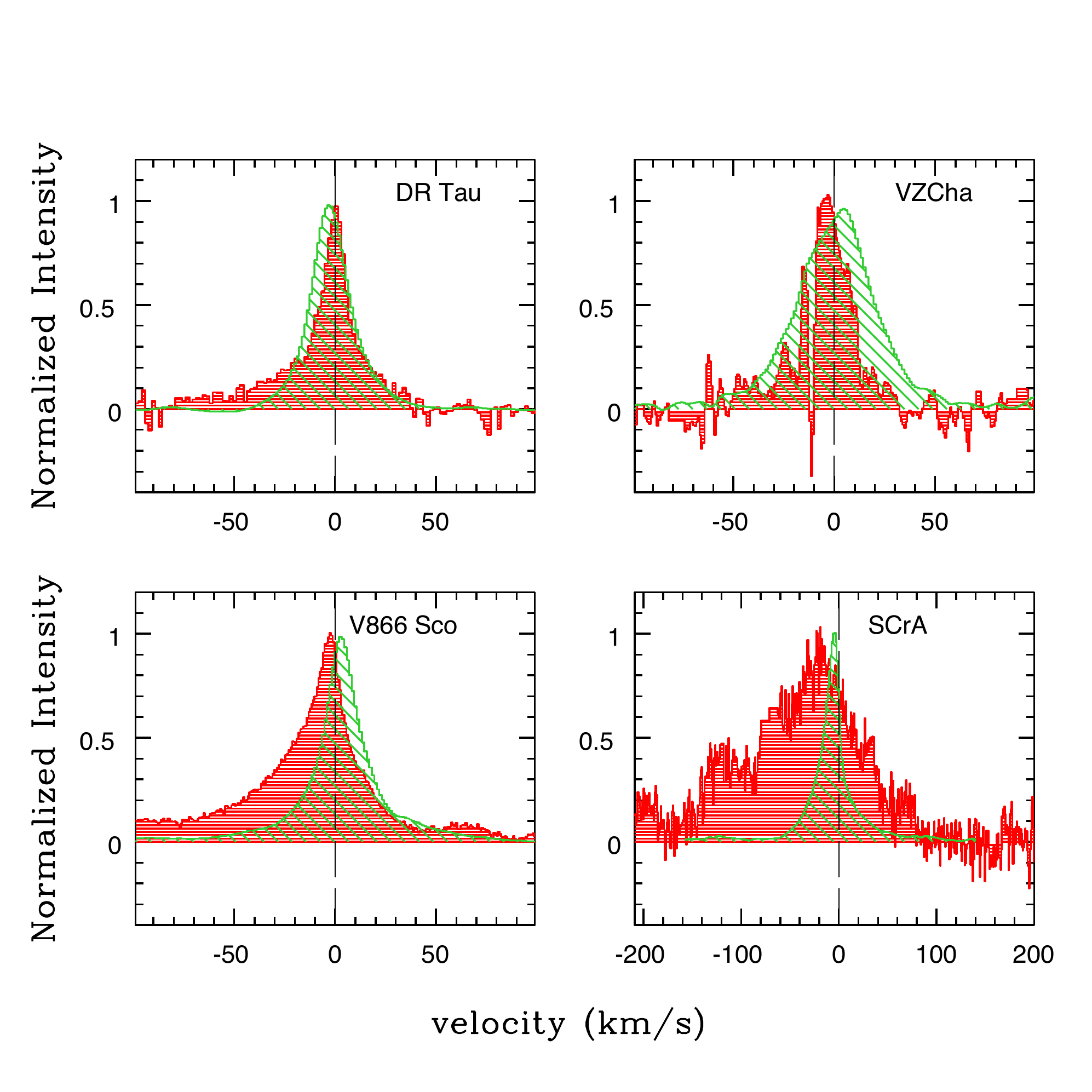}
\caption{{\small CO (green) -- {\oi}6300\AA\ (red) line profiles comparison. For SCrA~N (bottom right panel) the CO is compared with the {\oi}5577\AA\ line instead of the {\oi}6300\AA\ line.  }
 \label{CO_OI_comp}
}
\end{figure*}

Next, we compare the profiles of the strongest of the {\oi} lines, that at 6300\AA, with those of the {\neii} (tracing unbound gas)  and/or CO-rovibrational lines, in which the bulk component is mostly coming from the disk (Najita et al. 2003, Salyk et al. 2011), see  Figs.~\ref{Ne_CO_OI_comp}--\ref{Ne_OI_comp}--\ref{CO_OI_comp}.   
For this comparison we have degraded the spectra to the lowest resolution spectrum by Gaussian convolution and we have normalized the peak intensities to unity. 

Fig.~\ref{Ne_CO_OI_comp} shows the two Sample I objects  where all three gas diagnostics considered here are observed: TW~Hya and RU~Lup. 
The {\oi}6300\AA\ and the CO ro-vibrational profiles of TW~Hya are symmetric and centered at the stellar velocity, while the {\neii} line is blueshifted by $\sim 5$\,km/s and asymmetric (see Pascucci et al. 2011 for a detailed analysis). RU~Lup shows a complex {\oi} profile with blended HVC and LVC components, a CO single-peaked profile less blueshifted than the {\oi}LVC, and a {\neii} line dominated by the HVC (the LVC is likely not detected, see Sacco et al.~2012).

Fig.~\ref{Ne_OI_comp} shows the comparison between the {\oi}6300\AA\ and the {\neii} line profiles for  MP~Mus and V4046~Sgr. 
In both sources the {\oi}6300\AA\ line is broader than the {\neii} line and coincident with (in MP~Mus) or less blueshifted than (in V4046~Sgr) the {\neii} line.

Finally, Fig.~\ref{CO_OI_comp} shows the comparison of the oxygen and the CO ro-vibrational lines for the remaining four Sample I objects. We use as a reference transition the {\oi}6300\AA{} line for all sources except for SCrA~N for which we prefer to use the {\oi}5577\AA\ transition because it is less contaminated by the HVC. The {\oi} and CO profiles are quite different from each other and there is no trend in their FWHMs. For VZ~Cha, V866~Sco, and SCrA~N the CO line is less blueshifted than the {\oi}6300\AA\ line. Note that in the first two sources CO appears marginally {\em redshifted} with respect to the stellar velocity. For DR~Tau the {\oi} line is centered at the stellar velocity while the CO line may be slightly blueshifted: Bast et al.~(2011) report a peak centroid for the narrow component of -3.2$\pm$2\,km/s. 

In spite of the diversity in the line profiles, the following trends seem to emerge: i) the {\oi} LVC is broader and less blueshifted than the {\neii} line; ii) the {\oi} LVC is more blueshifted than the CO line; iii) their FWHMs do not follow a simple trend. 
These preliminary trends are consistent with a scenario in which the {\neii} line traces mostly the upper surface of a photoevaporative wind at larger radii than the CO and the {\oi} lines, which either probe bound gas in the disk or probe lower layers of the {\neii} wind. 
These trends are opposite to what is predicted by the models of Ercolano \& Owen (2010) where the {\oi} lines present a larger blueshift than {\neii} lines in primordial (not gapped) disks (see Sect.~4 for more details). 
In these models, the Ne$^+$ is produced by $\sim$1~keV X-rays in mostly neutral gas, and the {\neii} arises from somewhat cooler gas than the soft X-ray heated gas producing the {\oi}. Thus, the {\oi} emitting region is located higher up in the accelerating flow than the {\neii} emitting region. 
A larger sample size is definitively needed to confirm these trends and test this scenario. We will further discuss the {\oi} line profiles in Sects.~\ref{fwhm_peaks} and \ref{two_comp_sec}.

\subsection{Sample I and II}\label{sect_IandII}

To further investigate the origin of the {\oi} LVC we will consider the two samples collected here (the 8 TTs from the Sample I and the 30 TTs from the Sample II) in Sect.~\ref{correlations} and Sect.~\ref{ew_ratio_sect}. To analyze the kinematics of the {\oi} LVC (Sect.~\ref{fwhm_peaks}) we will use the whole Sample I and  a subsample of the Sample II, where the HVC and the LVC components can be well separated, as defined by HEG (see also Table~\ref{Sample_prop_tab}). 

\subsubsection{Correlations with stellar properties}\label{correlations}

\begin{figure}[!h]
\includegraphics[scale=0.25]{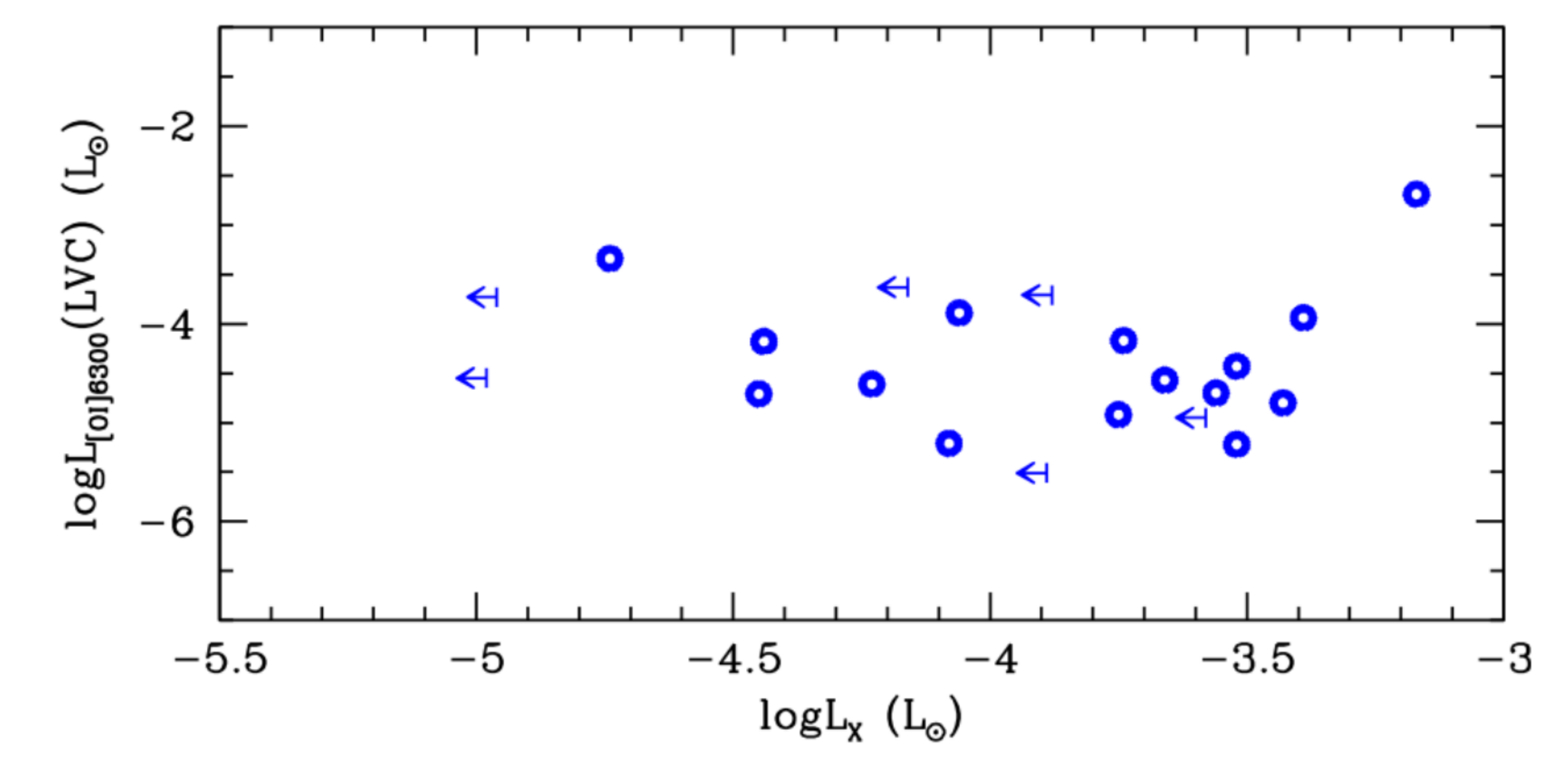}
\caption{{\small {\oi} luminosity versus X-ray luminosity for the subsample of 21 Sample II objects with X-ray luminosities (L$_X$) found in the literature. }
 \label{Lx_Lo}
}
\end{figure}

\begin{figure}[!h]
\includegraphics[scale=0.34]{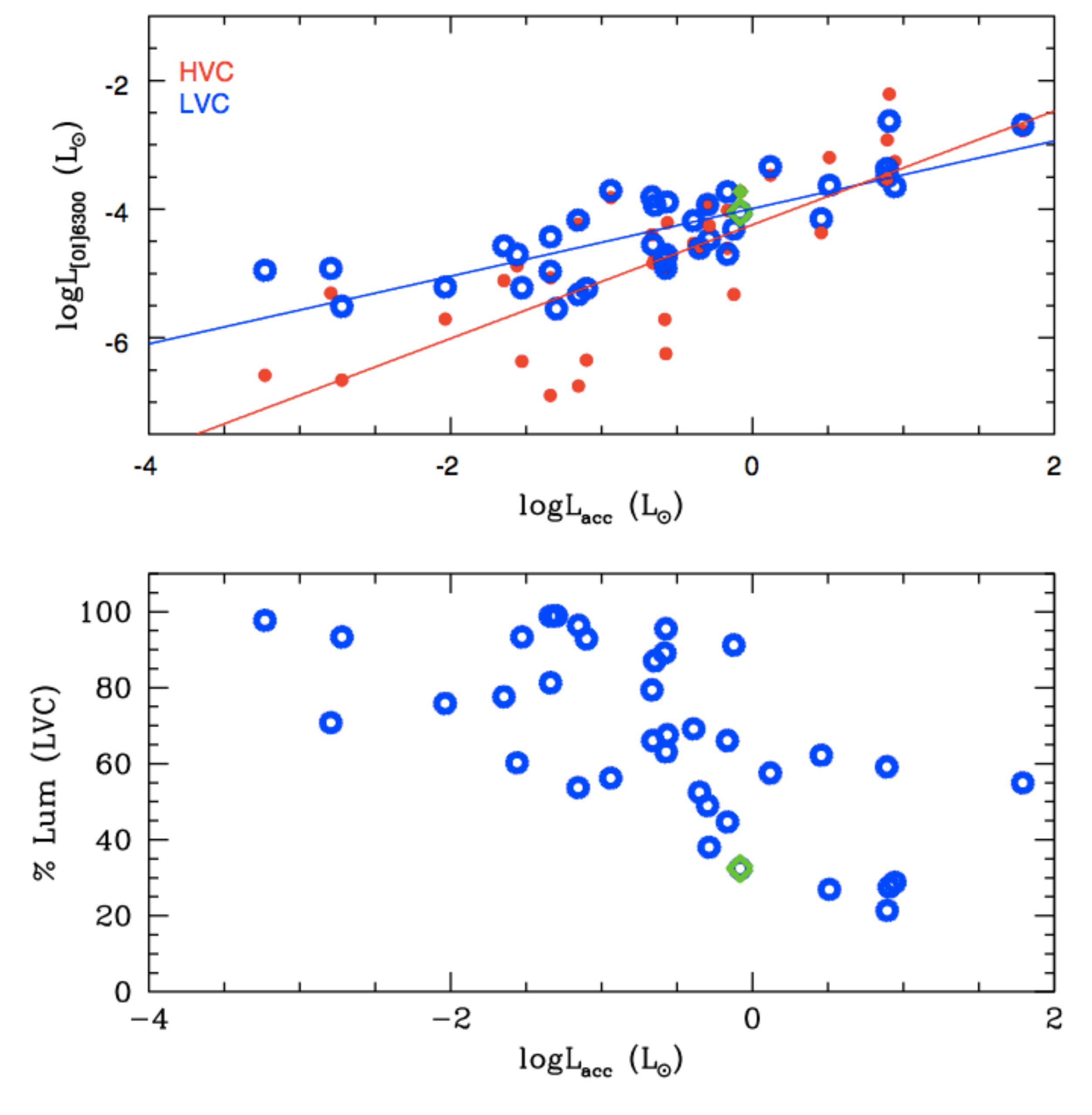}
\caption{{\small Top panel:  {\oi} luminosity versus \Lacc. Different colors refer to different components: 
blue open symbols refer to the LVC, red filled small circles refer to the HVC. 
Bottom panel: percentage of {\oi} LVC compared to the total {\oi} luminosity as a function of \Lacc. 
The green diamond represent S~CrA~N, which has not been considered in computing the relationship between $LogL_{{[O~{\sc I}]} LVC} - LogL_{acc}$ and $LogL_{{[O~{\sc I}]} HVC} - LogL_{acc}$ shown as blue and red solid lines respectively. 
}
 \label{lumOI_vs_Lacc}
}
\end{figure}

We have searched for correlations between the {\oi} LVC line luminosity ($L_{[O{\sc I}]LVC}$) and stellar properties for the complete Sample I and II objects (38 objects in total){\footnote {Note that S~CrA~N is marked as green diamond in the next figures and is not included in the correlations due to the large  contamination of the HVC to the LVC.}}.
We report no correlation between the $L_{[O{\sc I}]LVC}$ and the stellar luminosity or mass ($L_*$ and $M_*$ in Tables~2 and 3).
We also find no correlation between the $L_{[O{\sc I}]LVC}$ and the X-ray luminosity $L_X$ for the subsample of 21 Sample II objects (Stelzer et al. 2001, G\"udel et al. 2007, Fig.~\ref{Lx_Lo}). 
Instead, we identify a correlation between the $L_{[O{\sc I}]LVC}$ and the stellar accretion rate \Lacc\ (see Fig.~\ref{lumOI_vs_Lacc} top panel). We quantify this correlation by computing a linear least squares fit in $log$ scale: 
\begin{equation}
LogL_{[OI]LVC} = (0.52 \pm 0.07) \times LogL_{acc} - (3.99 \pm 0.09),
\label{lvc_lacc_eq}
\end{equation}

We note that the HVC, which is associated with jets/outflows, follows a different relation (see Figure~\ref{lumOI_vs_Lacc} top panel) and decreases steeply for low accretion luminosities. 
Figure~\ref{lumOI_vs_Lacc} bottom panel shows that the LVC dominates the {\oi} luminosity over a broad range of \Lacc\ after the accretion luminosity has decreased below $\sim$1~\Lsun. Even for high accretors the LVC contributes more than 20\% to the total {\oi} luminosity. A similar behavior was found by HEG when comparing average line profiles as a function of near-infrared colors, a proxy for mass accretion rates (see their Fig.~11).
  
Recently, Yang et al. (2012) found that \Lacc\  does not correlate with $L_X$ but correlates tightly with the total FUV luminosity ($L_{FUV}$). We have used this finding to convert our L$_{{[O~{\sc I}]} LVC}$~--~\Lacc\ relation into a L$_{{[O~{\sc I}]} LVC}$~--~L$_{FUV}$ relation:  
\begin{equation}
LogL_{[OI]LVC} = (0.63 \pm 0.09) \times LogL_{FUV} - (2.94 \pm 0.21)
\label{oi_fuv_eq}
\end{equation}
Higher accreting stars have higher FUV luminosities and higher {\oi} LVC luminosities. 
This positive correlation between stellar FUV luminosities and {\oi} luminosities hints to the {\oi} being a product of OH dissociation by FUV photons (Gorti et al. 2011).

The photodissociation of OH into hydrogen and oxygen atoms occurs when the molecule absorbs a photon with energy higher than 4.47~eV ($\lambda$=2616\AA, van Dishoeck \& Dalgarno 1983).  About 50\% of the time OH photodissociates to produce O in the $^1D$ state, and 5\% in the higher $^1S$ state (Festou \& Feldman 1981, Gorti et al. 2011 for a detailed explanation).
The $^1D-^3P$ transition produces the oxygen doublet at 6300\AA\ and 6363\AA, and the $^1S-^1D$ transition produces the emission at 5577\AA.

The photodissociaton of H$_2$O occurs via two main channels. In 79\% of the cases a water molecule dissociates into OH+H, and in the remaining 21\% of the cases O+2H are produced (Harich et al. 2000).
The latter case requires a threshold energy of the FUV photon of about 9.5~eV to produce oxygen in the electronic ground state.  The fraction of the oxygen atoms that are produced in the electronically excited D state is not known precisely, but Harich et al. (2000) argue that it is not a major channel even for energetically-allowed FUV photons with energies greater than about 11.5~eV.

Since most H$_2$O photodissociates to OH, and this is generally followed by OH photodissociation to oxygen, there are about equal numbers of H$_2$O and OH photodissociations.  
However, since H$_2$O only photodissociates to oxygen 21\% of the time and rarely produces oxygen in the $^1D$ state, the production of oxygen,  of O$^1D$, and therefore of {\oi}6300\AA\ are primarily by the photodissociation of OH.

We next make a simple calculation to show that the FUV luminosity is sufficient to produce the observed {\oi} via OH photodissociation, as suggested by the low 6300/5577\AA\ ratios described in the next section.  This calculation is done under the assumption that OH dominates the FUV opacity, and therefore absorbs all FUV photons that hit the disk. If $\dot N_{{OH}}$ is the number of FUV photons photodissociating an OH molecule per second, about half of these produce an O$^1$D atom in an excited state that then radiatively decays. Therefore, $L_{{[OI]}} \sim 1/2 \dot N_{{OH}}\times (hc/\lambda)$ where $hc/\lambda \sim$2eV is the energy of the 6300\AA\ photon. We make the assumption that Lyman $\alpha$ photons dominate the FUV flux (e.g. Bergin et al. 2003) and that a fraction $f_d$ of the stellar FUV flux is intercepted by the disk. We thus estimate that $\dot N_{{OH}} \sim f_d \times (L_{FUV}/10eV)$. Typically, $f_d \sim$ 0.5. We therefore find that L$_{{[OI]}} \sim$  0.05 L$_{FUV}$ is the maximum {\oi} luminosity that can be produced by OH if it  dominates the FUV opacity.  
This is consistent with the measured relation between L$_{{[OI]}}$ and  L$_{FUV}$, in other words the observed L$_{{[OI]}}$  are lower than the ones estimated in our simple calculation.

\subsubsection{{\oi} LVC lines ratio}
\label {ew_ratio_sect}

\begin{figure}[!h] 
\includegraphics[scale=0.29]{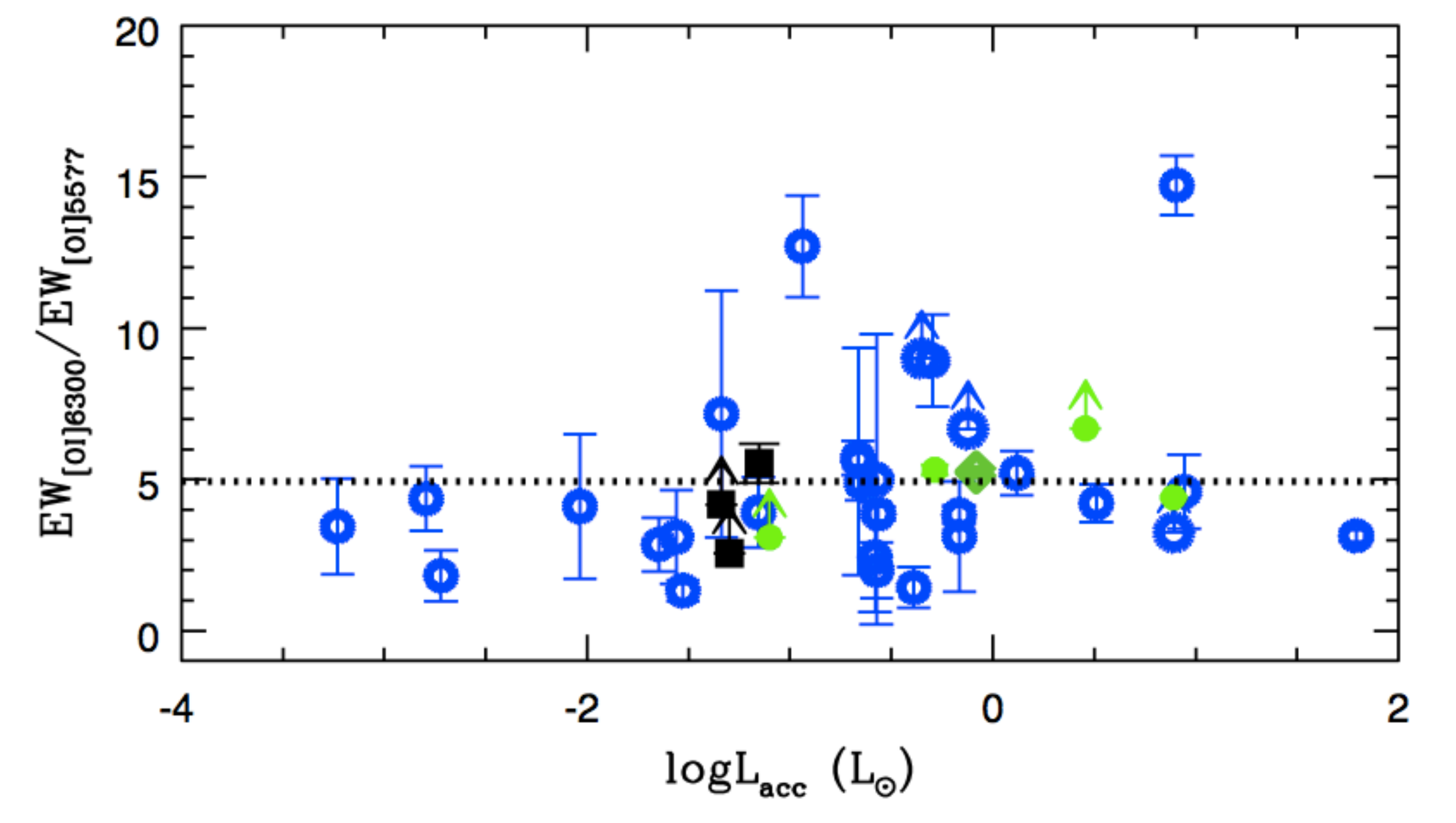}
\caption{ {\small 
EW {\oi}6300/5577\AA\ ratios versus Log\Lacc. Sample I objects are shown as black squares (objects with a {\neii} wind), and green filled circles (objects with a CO wind). The green diamond represents S~CrA~N. Sample II objects are shown  as blue open circles. The black dotted line represents the mean value measured  neglecting the lower limits. 
}
\label{EW_lacc_macc}
}
\end{figure}

As discussed in Sect.~\ref{intro}, the low {\oi} 6300/5577\AA\ line ratio of 7 found for TW~Hya (Pascucci et al. 2011) needs thermal emission coming from a very hot ($T \sim5,000-8,000$K) and dense 
($n_e \sim 10^7 - 10^8$cm$^{-3}$; $n_H \sim 10^8 - 10^9$cm$^{-3}$) gas (Kwan \& Tademaru 1995, Gorti et al. 2011). 
These physical conditions are unlikely in disks, especially in the photoevaporative flow. Such a low ratio can, however, be obtained from OH dissociation 
by FUV photons (Storzer \& Hollenbach 1998). OH photodissociation results in {\oi} 6300/5577\AA\ line ratios that can range from $\sim 7$ to lower values depending on the density of the molecular gas (e.g.  van Dishoeck \& Dalgarno 1984, Gorti et al. 2011). 
Here we show that low ratios are typical in TTs. 
Fig.~\ref{EW_lacc_macc} presents the {\oi} LVC 6300\AA/5577\AA\ EW ratios as a function of \Lacc{} for our Sample I and Sample II sources. 
We find that these ratios range between $\sim$1 and $\sim$8, with a mean value of $\sim$5, and change very little over six orders of magnitude in \Lacc.  There are two exceptions in our samples: DGTau and HNTau which have ratios of $\sim$15 and $\sim$13, respectively. Both sources power a strong jet (HEG and Hartigan et al. 2004) which makes it difficult to reliably estimate the EW of the LVC, especially in the 6300\AA\ line. The higher ratio may reflect a contamination of the HVC in the 6300\AA\ line profile, or indicate a higher contribution from thermal emission in these cases.  
There are 7 lower limits to the  6300\AA/5577\AA\ EW ratio due to non-detections of the {\oi}5577\AA\ line. These lower limits span over all the EW ratio covered by the detections and do not allow us to draw any further conclusion.

\subsubsection{FWHM and peak velocities of the LVC}
\label{fwhm_peaks}
FWHMs and peak velocities provide useful information about the gas kinematics. 
We list these quantities 
for our Sample I and for the subset of Sample II objects where the LVC and the HVC can be well separated in  Table~\ref{Sample_prop_tab}. As mentioned in Sect.~\ref{ew_section}, we have selected only Sample II objects with a LVC that is clearly distinct from the HVC (see HEG's Table~8 and Table~\ref{oi_Table} in this paper for the complete Sample I and II sources used in this analysis). As described in Sect.~\ref{sect_line_prof} we measure peak velocities and FWHMs to characterize, to a first order, the profiles of the {\oi} 6300\,\AA{}. However, for S~CrA~N, RU~Lup, and V866~Sco (marked with an asterisk in Table~\ref{oi_Table}) we opted to use the {\oi} 5577\,\AA{} line because the 6300\,\AA{} LVC is contaminated by the HVC (see Fig.~\ref{OIsameProfile}).

\begin{figure}[!h]
\includegraphics[scale=0.25]{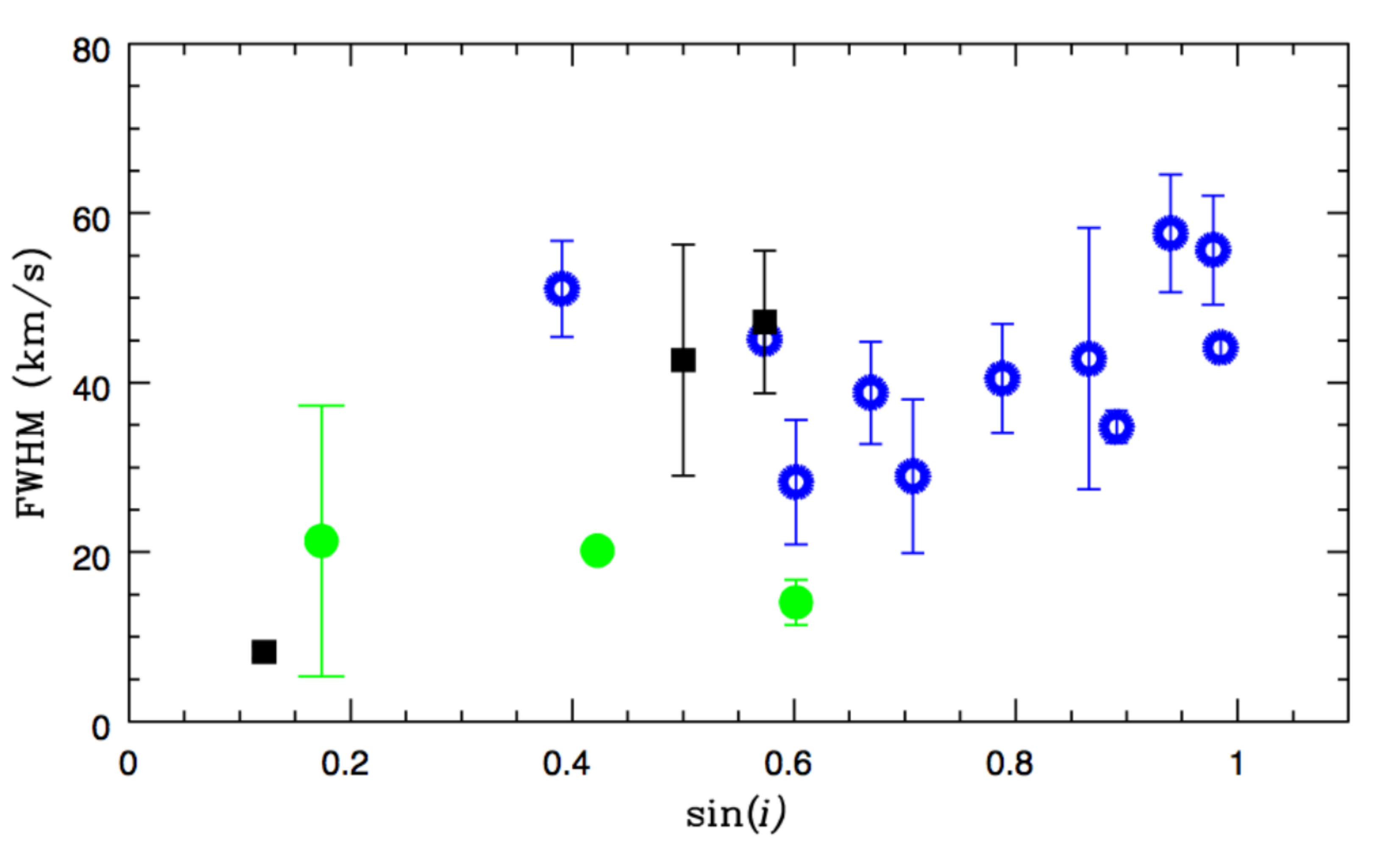}
\caption{
{\small  FWHM versus sine of the disk inclination. The blue open circles refer to a subsample of objects analyzed by HEG. The black squares refer to the Sample I objects with evidence of {\neii} emission. The green filled circles are for the Sample I objects with evidence of CO emission. }  \label{FWHM_vs_sini}
}
\end{figure}

Figure~\ref{FWHM_vs_sini} shows that the FWHMs range between 10--60~km/s (with the only exception of SCrA~N that exhibits a FWHM$\sim$110 km/s, see Tab.~\ref{Sample_prop_tab}), with several sources clustering around $\sim$40~km/s. 
The FWHM measured for the samples analyzed here  are larger than that predicted by models of EUV/X-ray photoevaporation (Ercolano \& Owen 2010). Moreover, the median FWHM suggests an origin in the inner part of the disk ($\sim$0.5~AU) under the assumption of Keplerian rotation.

In Tab.~\ref{Sample_prop_tab} we report the peak centroids (with respect to the star velocity). They span between +1~km/s and $\sim -$10~km/s, the mean being $-$5.0~km/s. 
The exception are SCrA~N for Sample I and YY~Ori and DQ~Tau for Sample II objects, where the largest blueshifts indicate a contamination from the HVC. 
Blueshifts in the {\oi} LVC were already noted by HEG and are a clear signature of unbound gas flowing outwards, the redshifted gas being occulted by the disk.  

Fig.~\ref{FWHM_vs_sini} shows a {\it possible} trend of increasing FWHM with increasing disk inclination. This can be explained by bound gas in Keplerian rotation around the star. 
At the same time the blueshifts in peak velocities should trace unbound gas.  
This suggests that the {\oi} LVC may itself have multiple components. This hint is corroborated by the analysis of the  highest resolution and S/N spectra (see Sect. 3.2.4).  
We have also searched for but not found any clear trend between the FWHM or peak velocity and the stellar accretion luminosity.

\subsubsection{Two components fit to the LVC}
\label{two_comp_sec}

\begin{figure}
\includegraphics[scale=0.43]{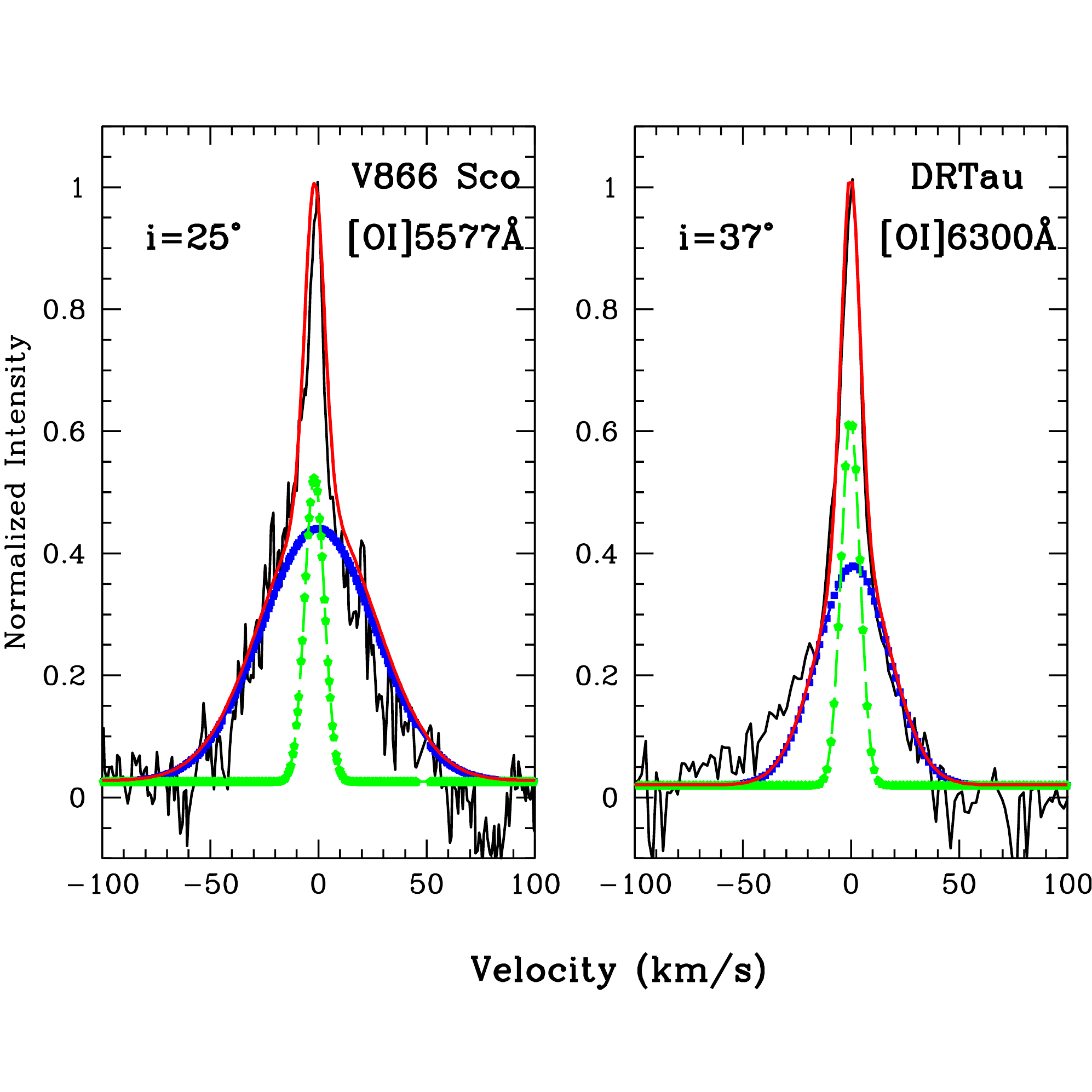}
\caption{{\small Double Gaussian fit for the DR~Tau and V866~Sco line profiles. Black: observed profile, red: summed profile, green: narrow-component LVC, blue: broad-component LVC.  }
\label{BC_NC}
}
\end{figure}

Here, we investigate the possibility that the {\oi} LVC can itself have multiple components. For this purpose we have chosen the two stars among the Sample I objects with S/N ratio around the {\oi} lines higher than 100, and with the higher resolution, namely V866~Sco (seen at at an inclination of 25$^\circ$) and DR~Tau (seen at 37$^\circ$). 
Figure~\ref{BC_NC} shows the 5577\AA\ line for V866~Sco and the {\oi} 6300\AA\ line for DR~Tau, both enlarged to show the $+/-$100~km/s velocity range. The first choice is motivated by the fact that a HVC is present in the {\oi} 6300\AA\ spectrum of V866~Sco (see Fig.~\ref{OIsameProfile}) and we want to minimize any contamination by the HVC in this analysis.
Unfortunately, the {\oi} 5577\,\AA{} line is not detected in the high-resolution UVES spectrum of DR~Tau so we analyze here the {\oi} 6300\,\AA{} profile.

The profiles in Fig.~\ref{BC_NC} show a broad-based emission and a narrow-peaked core. 
We therefore use two Gaussian profiles to fit the LVC: a narrow component (NC-LVC) to reproduce the central peaked emission and a broad component (BC-LVC) to account for the  broader wing emission. 

We define the range of velocities where the NC-LVC peaks and leave the gaussian parameters (namely amplitude, width, and peak centroid) free to vary. Then, we fit a gaussian profile over the total minus best fit NC-LVC profile. The total line profile (plotted in red in Fig.~\ref{BC_NC}) is the sum of the two gaussian profiles (green is the NC-LVC and blue is the BC-LVC).
The FWHMs and peak velocities for both components are reported in Table~\ref{BC_NC_tab}. 

We find that the spectrum of V866~Sco is well reproduced by two gaussians: a broad component  centered at $\sim$0\,km/s, and a narrow component blueshifted by $-$1.8\,km/s. The BC-LVC has a FWHM that is about a factor of 6 larger than that of the NC-LVC and contributes about half of the flux.

In the case of DR~Tau, our procedure finds that two gaussians centered around 0\,km/s can reproduce the total LVC observed profile. 
However, there is a residual weak but extended blue wing that is not fitted by the sum of the two gaussians.  
This excess emission could be a contamination from the HVC. More high-resolution spectra are needed to definitively address this point. 
For both objects the errors on the peak velocity are $\sim$0.7~km/s, as listed in Table~4. 

We find that the FWHM of the whole LVC is larger than that found for the NC-LVC.  For both sources the NC-LVC is reduced to $\sim$10~km/s, from 20 km/s for V866~Sco and 14 km/s for DR~Tau.

\begin{deluxetable}{c
        c c c c c} 
\tablewidth{0pt}
      
\tabletypesize{\scriptsize}
\tablecaption{Narrow
        component (NC) and broad component (BC) of the LVC
        \label{BC_NC_tab}}
\tablehead{
\colhead{Star}
        & \colhead{FWHM$^a$} & \colhead{$v_{peak}^a$} &
        \colhead{FWHM$^a$} & \colhead{$v_{peak}^a$} & $area$
        BC$^b$ \\
\colhead{}
        & \colhead{BC-LVC} & \colhead{BC-LVC} &
        \colhead{NC-LVC} & \colhead{NC-LVC} & (\%)\\
}
\startdata
DR Tau
        & 40.7 & -0.2 & 9.8 & -0.1 & 38\%\\
V866 Sco
        & 61.9 & -0.1 & 10.7 & -1.8 & 45\%\\
\enddata
\tablenotetext{}{$^a$
        FWHM and $v_{peak}$ are in km/s, $^b$ percentage of the area
        covered by the BC.  }
%\tablerefs{}
\end{deluxetable}

\section{Observational constraints on the origin of the {\oi} LVC}
\label{obs_constr}
Our observations and analysis provide four distinct constraints on the origin of the {\oi} LVC.

\begin{itemize} 
\setlength{\itemindent}{0cm}

\item[1.]{
Our comparison of {\neii}, {\oi}, and CO line profiles hints at a trend between their peak velocities, with the {\neii} line being more blueshifted than the {\oi} line which in turn is more blueshifted than the CO line. The average {\neii} peak velocity 
is $\sim -8$km/s (Sacco et al. 2012), the median of the {\oi} LVC is -5~km/s (this work and HEG), while the shifts in the CO line are $<5$\,km/s and may not be significant given the associated uncertainties (Bast et al. 2011). 
The X-EUV disk models of Ercolano \& Owen (2010) predict that {\oi} lines should have a larger blueshift than {\neii} lines in primordial (not gapped) disks. This is because in their models the {\oi} emitting region resides in hotter gas closer to the star and above  the {\neii} emitting region. 
As an example Ercolano \& Owen (2010) predict shifts of -6 and -3\,km/s for the {\oi} and {\neii} line centroids respectively for a disk inclined by 40$^\circ$. This trend is opposite to what we see in the data suggesting that the {\oi} emitting region is at larger disk radii or deeper in the disk than the {\neii} emitting region. The broader  {\oi} and CO profiles with respect to the {\neii} line suggest a larger contribution from bound inner disk gas for {\oi} and CO. 
}

\item[2.]{We measure {\oi} 6300\AA\ LVC line luminosities ranging from $10^{-3}-10^{-6}$\Lsun, 
with a mean value of $4.5 \times 10^{-5}$\Lsun. 
While these large values exclude emission from the EUV fully ionized layer, they are consistent with {\oi} tracing either thermal excitation in the soft X-ray-heated region (Hollenbach \& Gorti 2009, Ercolano \& Owen 2010) or non-thermal excitation in regions where OH is photodissociated by FUV photons into oxygen atoms  in excited electronic states (Gorti et al. 2011). However, the lack of correlation between the {\oi} luminosity and $L_X$  and the positive tight correlation with $L_{\rm acc}$ ($L_{\rm FUV}$) seems to favor the latter over the first scenario (Sect.~\ref{correlations}). We are cautious not to draw a stronger conclusion based solely on this observable because the measured $L_X$ may miss the soft-Xray component which, if present, may still irradiate the disk surface and heat it sufficiently to produce thermal {\oi} emission. 
Hints that evolved transitional disks may 'see' this soft X-ray component come from line ratios of ionic species detected at infrared wavelengths (Szulagyi et al. 2012). However, in classical disks, accreting at perhaps higher rates than transitional disks (Espaillat et al. 2012, Kim et al. 2013), the situation may be different.  
Strongly accreting sources are known to be accompanied by mass outflows (10-100 times lower than accretion rates, Cabrit et al. 1990, HEG), and Hollenbach \& Gorti (2009) calculate that soft X-rays will not penetrate the outflow column to irradiate the disk when \Macc\ is $\gtrsim 10^{-9}$ M$_\odot$/yr.
The majority of the stars analyzed here have $L_{\rm acc}>10^{-2}$\,$L_\odot$ (see Fig.~5), meaning \Macc $\gtrsim 10^{-9}$ M$_\odot$/yr. Therefore, outflows could absorb most of the soft X-ray component from the star. FUV photons, on the other hand, can penetrate the outflow column and reach the disk surface for \Macc $\lesssim 4\times 10^{-7}$ M$_\odot$/yr, basically for all but four stars in our sample.  

The FUV photon luminosity is high  enough for {\oi} photodissociation to explain the observed {\oi}-LVC luminosities. 
In the framework of the FUV model, the correlation shown in 
Eq.~\ref{oi_fuv_eq}  can be easily explained because more FUV stellar photons impinging on the disk imply more OH dissociation, hence higher {\oi} luminosities. }

\item[3.]{We measure {\oi}6300\AA/{\oi}5577\AA\ line ratios for a large sample of TTs and find values ranging from $\sim$1 to $\sim$8 over six orders of magnitude in $L_{\rm acc}$. 
These small values are difficult to produce in thermally heated gas present on the surface of protoplanetary disks but can be explained in a disk layer where {\oi} is produced by the non-thermal process of OH dissociation by FUV photons (Gorti et al. 2011). This observable independently points to {\oi} optical emission tracing stellar FUV photons impinging on the disk (see constraint 2).}

\item[4.]{The measured FWHM are larger than those predicted by the models accounting for disk photoevaporation. These larger FWHM suggest an origin of the line in the inner part of a disk in Keplerian rotation. Moreover
we find a possible trend of increasing FWHM of the {\oi} LVC with increasing disk inclination (Fig.~\ref{FWHM_vs_sini}). This correlation can be explained again by bound gas in Keplerian rotation around the star.  At the same time we confirm the average blueshift of $-$5~km/s found by HEG. 
This suggests the presence of multiple components within the {\oi} LVC. 
A detailed analysis of the {\oi} profiles for the two highest resolution and S/N spectra confirms this notion.  
The spectrum of V866~Sco is well reproduced by two gaussians: 
a broad component centered at $\sim$0~km/s, and a narrow component blueshifted by $\sim-$2~km/s. 
The BC-LVC could trace bound disk gas in Keplerian rotation around the star. Its broad FWHM of $\sim$60\,km/s points to {\oi} gas as close in as $\sim$0.2\,AU. If the BC-LVC profile lacks a double peak (which cannot be proven by these data due to contamination of the NC-LVC) then gas should extend out to $\sim$10\,AU\footnote{We also tested Keplerian profiles for the BC-LVC and found that double-peaked Keplerian profiles resulted in less good fits than single-peaked Keplerian profiles}.
The BC-LVC contributes to about 40\% on the total flux (Tab.~\ref{BC_NC_tab}).  
In the case of DR~Tau, the observed spectrum is reproduced by two gaussians centered around 0\,km/s , even if this configuration cannot reproduce the excess emission on the blue side of the {\oi} 6300\AA{} profile. This excess emission contributes to less than 10\% on the total {\oi} LVC emission and could either be a wind component or contamination by the HVC. 
The FWHM of the BC-LVC is $\sim$40~km/s, and points to {\oi} gas as close in as $\sim$0.6\,AU.
In both the sources, the FWHM of the NC-LVC is $\sim$10~km/s. 
This component may be associated with that portion of the {\oi} gas that becomes gravitationally unbound at radial distances $\ge10$\,AU. 
For gas heated to T$\sim$1,000~K by FUV or hard X-ray photons, the sound speed is $\sim$2~km/s, 
corresponding to the blueshift found for the NC-LVC of V866~Sco. 
The average blueshift found for the sample analyzed here and by HEG is larger, $\sim$5~km/s. 
Our blueshifts may be consistent with a mainly molecular photoevaporative wind, but given the uncertainties, the origin of the NC-LVC cannot be pinned down. 
}

\end{itemize}

The ensemble of data presented here supports the following scenario for the origin of the {\oi} LVC. Stellar FUV photons penetrate the outflow columns of classical T Tauri stars and reach the disk surface where they dissociate OH molecules resulting in oxygen atoms in the $^1D$ and $^1S$ states. The atoms then decay to the ground state via the {\oi} 6300 and 5577\,\AA{} lines. 
Higher accretion rates imply higher FUV luminosities, higher rates of OH dissociation, and hence higher {\oi} luminosities as observed. 
This photodissociated layer may have a bound component of gas in Keplerian rotation and unbound gas at radial distances $\ge10$\,AU, likely a photoevaporative wind. This wind is heated by either the same FUV photons that dissociate OH molecules, or by hard ($>1$ keV) X-ray photons that can also penetrate the outflow column. These hard X-ray photons may be responsible for producing the {\neii} blueshifted emission from gas at higher vertical heights of the accelerating photoevaporative flow.

Distinguishing the bound from the unbound component and determining the wind-driving agent is key to measuring photoevaporative mass loss rates. If FUV-driven, one might expect that the peak of the flowing component (NC-LVC) correlates with the FUV luminosity 
(more FUV would lead to higher temperature, and the higher the temperature, the faster the flow, hence the larger the blueshifts). 
In this case the FUV radiation is not only photodissociating the OH molecules, but is also driving the flow. On the other hand, if the flow is X-ray driven, one might expect a correlation between the peak centroids of the NC-LVC and the X-ray luminosity. In this case, the {\oi} is produced by the FUV photodissociation of the OH molecules, but X-ray photons heat and launch this molecular flow. Since the hard X-ray flux is well determined, we searched for trends between $v_{peak}$ and L$_X$ for our sample. We found no correlation with L$_X$ (correlation coefficient $\sim$0.01) and only a weak correlation with L$_{FUV}$ (correlation coefficient $\sim$-0.33), and therefore cannot make any strong conclusions. A larger sample of high resolution spectra is required to decide between the two mechanisms driving the wind.

As disks evolve and mass accretion rates decline below \Macc $\lesssim 10^{-9}$ M$_\odot$/yr, EUV (and soft X-rays) can start penetrating the outflow columns, and EUV photons rather than X-rays may be responsible for producing {\neii} emission in these disks. The lack of blueshift in the {\oi} line from the evolved disk of TW Hya (Pascucci et al. 2011) is consistent with our scenario. In this case the bound {\oi} component dominates and there is no detectable unbound component. Given the lower mass accretion rate of this star (lower FUV photon flux) and the less flared disk surface with respect to that of classical T Tauri disks (Gorti et al. 2011), FUV or X-rays may be inefficient in driving a dense molecular photoevaporative wind. Extrapolating our scenario to disks around stars with higher mass accretion rates than classical TTs, we should expect a decline in the {\oi} LVC as fewer stellar FUV photons can penetrate more massive winds.  Indeed, for the Class I/Class II stars in Taurus-Auriga from the sample of White \& Hillenbrand (2004) we note that while the  {\oi} LVC in their optical spectra is always detected in Class II objects, it is missing in some Class I objects (e.g. L1551-IRS5, HLTau).

\section{Summary}

To understand the origin of the {\oi} LVC we analyzed archival high-resolution optical spectra for two samples of objects: {\it i)} a sample of stars with previously detected {\neii}$\lambda$12.81$\mu$m and/or CO ro-vibrational 4.7$\mu$m emission and with  profiles suggestive of a slow disk wind; and {\it ii)} the comprehensive survey of classical TTs in Taurus collected by HEG.  

Our comparison of {\oi} LVC profiles with the {\neii} and CO profiles identifies a possible trend in peak centroids with the {\neii} being more blueshifted than the {\oi}, which, in turn, is more blueshifted than the CO line. In addition, the broader profiles of the {\oi} and CO with respect to the {\neii} line point to the presence of more bound gas contributing to the {\oi} and CO emission. These results are consistent with the three gas diagnostics  
tracing different parts of a photoevaporative flow, with the {\neii} being likely in the top layer.

We discover a correlation between the luminosity of the {\oi} LVC  and stellar accretion rates ($L_{\rm acc}$), which we consistently compute for the Sample I and II objects. Because $L_{\rm acc}$ is known to positively correlate with the stellar FUV luminosity, our result implies that the higher the stellar FUV luminosity the higher the {\oi} luminosity of the LVC. 
The correlation with ($L_{\rm acc}$) and the {\oi} luminosity of the LVC implies that if there is an unbound component  to the LVC from a photoevaporative flow, the flow is stronger for stars with higher accretion rates. This is in contrast to the usual assumption that photoevaporation is only important in late stage disk evolution. Instead, it may be proceeding throughout the T Tauri phase, at rates that decrease as the accretion rate decreases.

We also show that the {\oi}  6300/5577\AA\ EW ratios span only a narrow and low range of values (1-8) over six orders of magnitude in $L_{\rm acc}$ and most likely are a result of OH photodissociation. 
 We measure larger FWHM than those predicted by the photoevaporating models, and find a trend of increasing FWHM with increasing disk inclinations for the {\oi} LVC which points to bound gas in Keplerian rotation around the star. 
At the same time the average $-5$km/s peak velocity points to an unbound gas component. 

The analysis of the two highest resolution and signal-to-noise spectra shows that indeed the {\oi} LVC traces at least two different components: bound disk gas in Keplerian rotation and unbound gas farther out ($\ge 10$\,AU).
Higher resolution spectra on a larger sample of disks are necessary to test our inference of multiple LVC components and their ubiquity. 

Our results demonstrate that a thermal origin of the {\oi}  in a X-ray-heated photoevaporative wind, as proposed by Ercolano \& Owen (2010),  has several shortcomings. It cannot explain the trend in peak velocities between {\neii} and {\oi} lines and cannot easily account for: i) the lack of correlation between the   
{\oi} luminosity and $L_X$, ii) the correlation of $L_{[OI]}$ and $L_{\rm FUV}$ extending beyond 
mass flow rates that should block soft X-ray photons, and iii) the low {\oi} 6300/5577\AA\ line ratios. 
Based on our analysis, we propose an alternative scenario in which the {\oi} LVC traces the disk layer where stellar FUV photons dissociate OH molecules in oxygen and hydrogen atoms. Part of this layer may  participate in a  photoevaporative wind that could be driven either by FUV or by hard X-ray photons.  

A larger sample of high resolution spectra is definitively required to decide between the two mechanisms driving the wind.

\acknowledgments

The authors thank Jeanette Bast for providing us the CO line profiles, and Germano Sacco  for providing us the [NeII] line profiles. We also thank Barbara Ercolano and James Owen for useful discussions about the X-ray photoevaporation.
Support for E.R. was provided by Astronomy and Astrophysics research
grant to I.P. (ID:AST0908479). U.G. acknowledges support from a NASA ADAP grant (NNX09AC78G).  
The authors thank the anonymous referee for helpful comments that helped improving the clarity of this paper.

\appendix

\begin{figure*}
\centering
\includegraphics[scale=0.48]{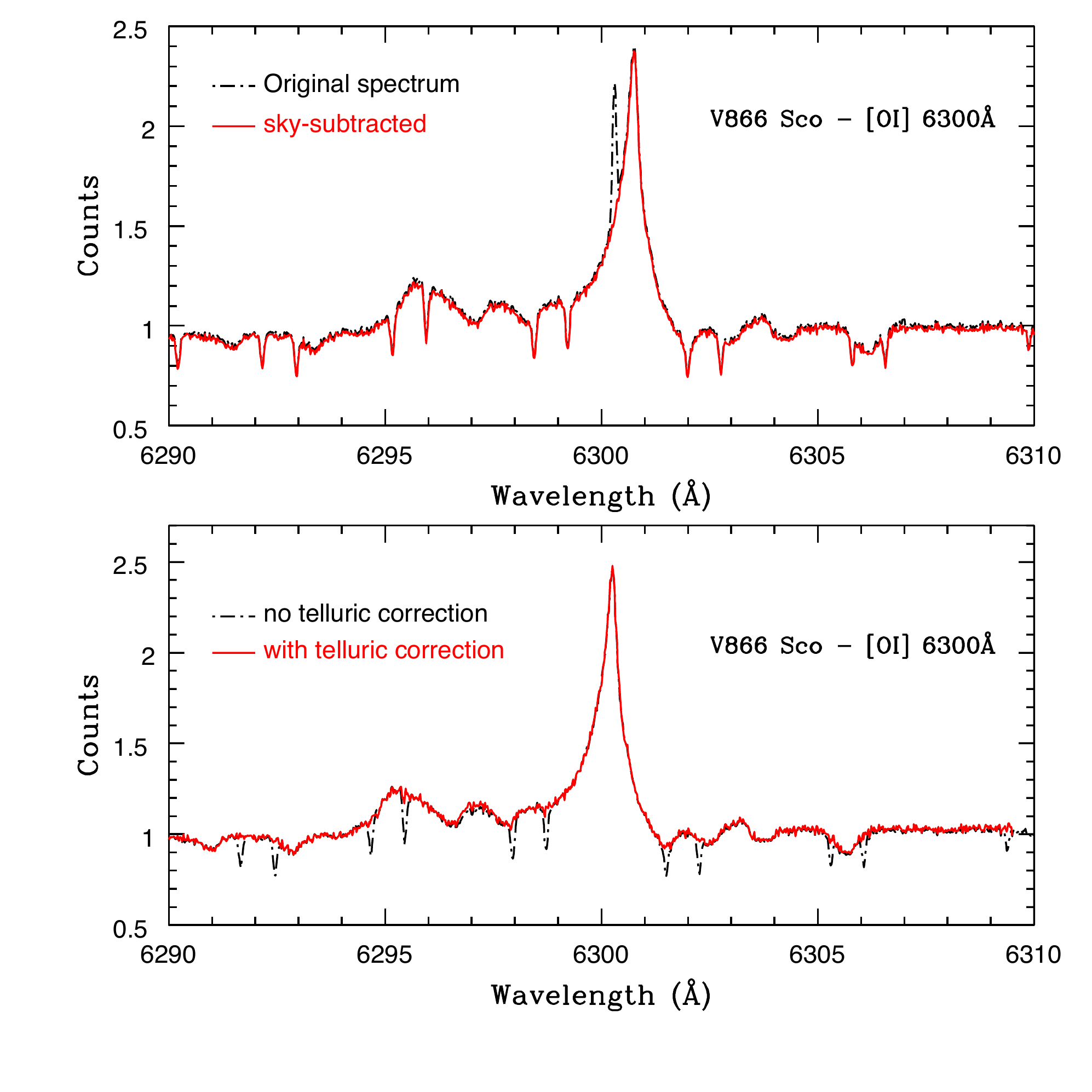}	
\caption{
{\small V866 Sco in the region around the {\oi}$\lambda$6300\AA\ line. Top panel: original spectrum (dashed, black profile) and sky-subtracted spectrum (solid, red profile). The profiles are corrected for the heliocentric velocity but not for the radial velocity of the star. Bottom panel: spectrum of the target star with no correction for the telluric features (dashed, black profile) and after telluric correction (solid, red line). The profiles in this panel are in the stellocentric reference frame. All the spectra presented in this paper have been corrected for terrestrial emission and absorption features.}  \label{v866_redu}
}
\end{figure*}

We report in Fig.~\ref{v866_redu} the reduction steps applied to the spectrum of V866~Sco.

\end{document}